\documentclass[showpacs,aps,prd,reprint,superscriptaddress,nofootinbib,longbibliography,preprintnumbers]{revtex4-1}
\usepackage[dvipdfmx]{graphicx}
\usepackage{graphicx,bm,color}
\usepackage{subfigure}
\usepackage{amsmath}
\usepackage{amssymb}
\usepackage{amsfonts}
\usepackage{cases}
\usepackage{cancel}
\usepackage{hyperref}
\usepackage{ulem}
\usepackage{amsmath,amssymb,bm,color,longtable,mathrsfs,amsfonts,slashed}

\usepackage{here}
\usepackage{tikz}
\usetikzlibrary{matrix}

\begin{document}

\title{
Sound velocity peak induced by the chiral partner in dense two-color QCD}

\author{Mamiya Kawaguchi}
\email[]{mamiya@ucas.ac.cn}
\affiliation{ 
School of Nuclear Science and Technology, University of Chinese Academy of Sciences, Beijing 100049, China
} 
\author{Daiki~Suenaga}
\email[]{daiki.suenaga@kmi.nagoya-u.ac.jp}
\affiliation{Kobayashi-Maskawa Institute for the Origin of Particles and the Universe, Nagoya University, Nagoya, 464-8602, Japan}
\affiliation{Research Center for Nuclear Physics, Osaka University, Ibaraki 567-0048, Japan}

\begin{abstract}
Recently, the peak structure of the sound velocity was observed in the lattice simulation of two-color and two-flavor QCD at the finite quark chemical potential. The comparison with the chiral perturbation theory (ChPT) result was undertaken; however, the ChPT failed in reproducing the peak structure.
In this study, to extend the ChPT framework,
we incorporate contributions of the $\sigma$ meson, that is identified as the chiral partner of pions, on top of the low-energy pion dynamics by using the linear sigma model (LSM). Based on the LSM we derive analytic expressions of the thermodynamic quantities as well as the sound velocity within a mean-field approximation. As a result, we find that those quantities are provided by sums of the ChPT results and corrections, where the latter is characterized by a mass difference between the chiral partners, the $\sigma$ meson and pion. 
The chiral partner contributions are found to yield a peak in the sound velocity successfully. We furthermore show that the sound velocity peak emerges only when $m_\sigma >
\sqrt{3}m_\pi$ and $\mu_q >  m_\pi$, with $m_{\sigma(\pi)}$ and $\mu_q$ being the $\sigma$ meson (pion) mass and the quark chemical potential, respectively. The correlation between the sound velocity peak and the sign of the trace anomaly is also addressed.
\end{abstract}

\maketitle

\section{Introduction}

The sound velocity is a crucial ingredient for characterizing the baryonic dense matter, particularly in the context of neutron stars. Recent discussions have emphasized that the presence of a peak structure in the sound velocity is an essential feature for meeting the observation of the mass and radius relation in neutron stars~\cite{Baym:2017whm,Huang:2022mqp}.
To clarify the presence of the sound velocity peak from the first-principles viewpoint of quantum chromodynamics (QCD), there is a growing demand for lattice QCD simulations including the quark chemical potential ($\mu_q$). 
However, it is not easy to accomplish lattice simulations in high dense regions, primarily due to the so-called {\it sign problem} in three-color QCD at finite $\mu_q$~\cite{Aarts:2015tyj,Nagata:2021ugx}.

To avoid the sign problem, 
QCD-like theories are considered instead of the real-life QCD with three colors.  
For instance, the sign problem is absent in the case of two-color QCD (QC$_2$D) with two flavors, owing to the pseudoreality of $SU(2)_c$ gauge group~\cite{Muroya:2003qs}. 
This advantageous feature enables the implementation of lattice simulations in large $\mu_q$ regions in QC$_2$D. 

In QC$_2$D, the diquarks, made of two quarks, form color-singlet baryons due to again the pseudoreality of $SU(2)_c$, and thus they obey the Bose-Einstein statistics. When the chemical potential $\mu_q$ reaches half the value of the pion mass, the ground-state diquark baryons become massless.  
Consequently, this gives rise to the Bose-Einstein condensation of diquarks, i.e., the {\it diquark condensation}, along with the creation of a baryonic matter, resulting in a transition to the {\it baryon superfluid phase}~\cite{Kogut:1999iv,Kogut:2000ek}. 
In contrast to the large $\mu_q$,
the {\it hadronic phase} is realized in the lower $\mu_q$, which is smoothly connected to the vacuum. 
Lattice simulations have been extensively performed to reveal various aspects in the cold dense QC$_2$D such as phase structures, thermodynamics quantities and the hadron mass spectrum, across the phase transition~\cite{Hands:1999md,Kogut:2001na,Hands:2001ee,Muroya:2002ry,Chandrasekharan:2006tz,Hands:2006ve,Alles:2006ea,Hands:2007uc,Hands:2010gd,Hands:2011hd,Cotter:2012mb,Hands:2012yy,Boz:2013rca,Braguta:2016cpw,Boz:2018crd,Astrakhantsev:2018uzd,Iida:2019rah,Wilhelm:2019fvp,Boz:2019enj,Iida:2020emi,Astrakhantsev:2020tdl,Bornyakov:2020kyz,Murakami:2022lmq}
(for a review, see~\cite{Braguta:2023yhd}).

Recently, the $\mu_q$ dependence of the sound velocity has been evaluated in the QC$_2$D lattice simulation~\cite{Iida:2022hyy}. The simulated result has indicated that in the baryon superfluid phase, the sound velocity monotonically grows from zero as $\mu_q$ increases, and soon it exceeds the conformal limit $c_s^2=1/3$. Meanwhile, at sufficiently large $\mu_q$ the relevant scale of the dense matter is solely provided by $\mu_q$, resulting in that the sound velocity eventually converges on $c_s^2=1/3$.
Therefore, the observed excess implies the appearance of peak structures in the sound velocity. 
Note that the isospin chemical potential dependence of 
the sound velocity in the three-color QCD has also been investigated through lattice simulations~\cite{Detmold:2012wc,Brandt:2018bwq,Brandt:2022hwy,Abbott:2023coj,Hippert:2024hum}, implying the presence of the sound velocity peak in the isospin asymmetric matter of the real-life QCD.

The high-density behavior of the sound velocity has also been theoretically under investigation in both the real-life QCD and QCD-like theories, based on quark-level pictures and their extensions~\cite{Kojo:2021wax,Fujimoto:2020tjc,Braun:2022jme}  including the quarkyonic description~\cite{McLerran:2018hbz}. Besides, impacts of the quark saturation inside hadrons on the sound velocity peak have also been examined~\cite{Kojo:2021ugu,Kojo:2021hqh,Chiba:2023ftg}, while considering the relation to the continuous transition from hadron to quark matter~\cite{Masuda:2012ed,McLerran:2018hbz}. Despite those extensive efforts, no conclusive picture for the peak structure of the sound velocity has been established yet.

In QC$_2$D with large $\mu_q$, the gluon is unaffected by the Meissner effect~\cite{Kojo:2014vja,Suenaga:2019jjv,Kojo:2021knn} since the diquark condensate exists as a color singlet object.
This leads to the gluon sector in QC$_2$D being unscreened and implies that the highly dense QC$_2$D system is expected to remain confined~\cite{Son:2000by,Son:2000xc,Kanazawa:2009ks}.
Indeed, the recent lattice simulation has shown that the Polyakov loop $L$ maintains small values even after the phase transition to the baryon superfluid phase: $L\ll1$~\cite{Boz:2019enj,Iida:2019rah}.
This indicates that QC$_2$D would be in the confinement phase in a wide range of $\mu_q$ and the hadronic picture could be applicable across the baryon superfluid phase transition, although at the extremely high-density region the quark degrees of freedom become significant. The difficulty in clarifying the microscopic pictures of cold dense matter is related to the quark-hadron continuity~\cite{Baym:2017whm}.
In this study, we approach the baryonic matter from the low-energy perspective using hadronic models. 
These models allow us to continuously
enter the baryon superfluid phase based on the hadronic picture, which would be one of the reliable methods to elucidate properties of the sound velocity from the low-energy viewpoint of dense QC$_2$D.


As a conventional approach of such a hadron effective model,
the chiral perturbation theory (ChPT) is frequently employed to make a comparison with lattice observations of cold dense QC$_2$D~\cite{Kogut:1999iv,Kogut:2000ek,Lenaghan:2001sd,Metlitski:2005db,Kanazawa:2009ks,Harada:2010vy}. 
In particular, the ChPT evaluation for the sound velocity within the mean-field approximation exhibits no explicit model parameter dependence~\cite{Son:2000by,Hands:2006ve}, and for this reason, the ChPT results serve as a robust benchmark of the low-energy QC$_2$D in comparing with the lattice data. Indeed, the ChPT prediction of the sound velocity is in good agreement with the lattice data in the vicinity of the phase-transition point; however, it monotonically converges to $c_s^2=1$ at high density without showing any peak structures. Hence, there is a discrepancy between the QC$_2$D lattice observation and the ChPT prediction of slightly higher $\mu_q$. 
This is because the ChPT is constructed upon a low-energy expansion with respect to the Nambu-Goldstone (NG) bosons, and this model is adaptable for only the deep low-energy regime of QC$_2$D. Thus, it is natural that the ChPT cannot apply to the high-dense regime, owing to the lack of higher excitations. 

Motivated by the shortcomings of the ChPT, in the present study we incorporate overlooked contributions from excited hadrons into an effective model, in order to bridge the gap between the QC$_2$D lattice observation and the ChPT evaluation.
In particular, we make use of the linear sigma model (LSM) invented in Ref.~\cite{Suenaga:2022uqn} to take into account additional contributions of the chiral partner, i.e., the $\sigma$ meson,
which is linked with the NG bosons under the chiral symmetry. 
Then, we discuss how the $\sigma$ meson contributes to the sound velocity to generate the peak structure. Besides, the recent QC$_2$D lattice simulation provides the $\mu_q$ dependence of the trace anomaly, indicating that the trace anomaly becomes negative in high-density regions~\cite{Iida:2022hyy}. We further examine the correlation between the sound velocity and the trace anomaly within the LSM.

This paper is organized as follows. In Sec.~\ref{sec:ChPTReview} we provide a brief review of the ChPT approach at a mean-field level, and present analytic expressions of the thermodynamics quantities including the sound velocity. Those quantities are also evaluated within the LSM approach in Sec.~\ref{sec:LSMFramework}, and the $\sigma$ meson contributions on top of the ChPT framework is clarified in this section. Then, in Sec.~\ref{sec:SoundVelocityLSM}, numerical demonstrations to visualize the $\sigma$ meson contributions are provided, and finally, in Sec.~\ref{sec:Conclusions}, we conclude the present work.

\section{ChPT in cold dense QC$_2$D}
\label{sec:ChPTReview}

In this study,
we aim to delve into the understanding of the sound velocity from a perspective of the hadronic picture. In particular, we take into account hadron contributions absent in the ChPT framework and extend the hadronic result of the sound velocity beyond the low-energy regime governed by the ChPT.
To facilitate our discussion, in this section, we provide a brief review of the ChPT which serves as a low-energy effective model for the lightest hadrons identified as the NG bosons associated with the chiral symmetry breaking.
We also provide an overview of the well-established thermodynamic quantities in the ChPT. Following that, we review the evaluation of the sound velocity in finite quark chemical potentials at zero temperature.

\subsection{ChPT Lagrangian}

In this subsection, we introduce the ChPT Lagrangian for two-flavor QC$_2$D which can access the baryonic matter.

In QC$_2$D, two (anti)quarks can be bound together by the strong interaction to form color-singlet (anti)baryons in addition to mesons, carrying the quark number $+2$ ($-2$). 
In contrast to the three-color QCD, those diquark baryons behave as bosonic particles sharing common properties with mesons due to the pseudoreal nature of color $SU(2)_c$ group. Hence, 
both the diquark baryons and mesons are treated on an equal footing in constructing the hadron effective model in QC$_2$D. Besides, it is known that the pseudoreality allows us to extend $SU(2)_L\times SU(2)_R$ chiral symmetry to the Pauli-G\"ursey $SU(4)$ symmetry, and in the low-energy region, this $SU(4)$ symmetry is spontaneously broken to the $Sp(4)$ symmetry~\cite{Kogut:1999iv,Kogut:2000ek}. Therefore, in QC$_2$D, low-energy effective models are constructed upon the symmetry-breaking pattern of $G=SU(4) \to H=Sp(4)$.

To formulate the ChPT Lagrangian, we introduce the nonlinear field $\xi$ which is parametrized by the NG bosons $\pi^i$,
\begin{eqnarray}
\xi&=&\exp\left(
\frac{i\pi^i X^i}{f_\pi}
\right). \label{XiDef}
\end{eqnarray}
Here, in two-flavor QC$_2$D, the manifold of $G/H=SU(4)/Sp(4)$ is characterized by five degrees of freedom, and hence the superscript ``$i$'' in Eq.~(\ref{XiDef}) runs from $1$ to $5$. In terms of the hadrons, $\pi^{i=1,2,3}$ and $\pi^{i=4,5}$ denote the pseudoscalars and (anti)baryons serving as the NG bosons, i.e., pions and positive-parity (anti)diquarks. The $4\times4$ matrices $X^i$ are generators belonging to the Lie algebra of $G/H=SU(4)/Sp(4)$ given by
\begin{eqnarray}
X^i&=&
\frac{1}{2\sqrt{2}}
\begin{pmatrix}
\tau_f^i&0\\
0& (\tau_f^i)^T
\end{pmatrix}
\;\;\; (\mbox{for $i=1,2,3$}),
\nonumber\\
X^i&=&
\frac{1}{2\sqrt{2}}
\begin{pmatrix}
0& D_f^i\\
(D_f^i)^\dagger& 0
\end{pmatrix}
\;\;\; (\mbox{for $i=4,5$}) , \label{GeneratorsX}
\end{eqnarray}
with $\tau_f^i$ being the Pauli matrix in the flavor space, and $D_f$ being $D_f^4= \tau_f^2$ and $D_f^5=i\tau_f^2$. The constant $f_\pi$ in Eq.~(\ref{XiDef}) is the pion decay constant at the vacuum. Since the nonlinear field $\xi$ is the representative of the coset space $G/H$, its transformation law can be determined by
\begin{eqnarray}
\xi \to g \xi h^T, \label{XiTransform}
\end{eqnarray}
with $g\in G$ and $h\in H$. Since $h$ satisfies 
\begin{eqnarray}
h^TE h=E\;\;\;
\mbox{with}\;\;\;
E=
\begin{pmatrix}
0&{\bm 1}\\
-{\bm 1}&0
\end{pmatrix} 
\end{eqnarray}
from its definition, one can define a useful chiral field $U$ which transforms homogeneously under the transformation~(\ref{XiTransform}) as
\begin{eqnarray}
U&=&\xi E^T \xi^T .
\end{eqnarray}
Indeed, this $U$ exhibits the following transformation law:
\begin{eqnarray}
U&\to& gU g^T.
\end{eqnarray}
Using the chiral field $U$, one can write down the $SU(4)$-invariant ChPT
Lagrangian of the lowest order in terms of the derivative (momentum) expansion~\cite{Kogut:1999iv,Kogut:2000ek}. :
\begin{eqnarray}
{\cal L}_{\rm ChPT}
&=& \frac{f_\pi^2}{4}{\rm tr}\left[  D_\mu U^\dagger D^\mu U  \right]
 \nonumber\\
&& +\frac{f_\pi^2m_\pi^2}{4} {\rm tr}\left[EU + U^\dagger E^\dagger    \right],
\label{ChPT_Lag}
\end{eqnarray}
where
$m_\pi$ denotes the pion mass at the vacuum.
The first term in Eq.~(\ref{ChPT_Lag})
is the kinetic term of the chiral field $U$ and
the quark number chemical potential $\mu_q$ is embedded in the covariant derivative,
\begin{eqnarray}
D_\mu U = \partial_\mu U - i \mu_q \delta_{\mu0} (JU +U J^T) ,
\end{eqnarray}
with
\begin{eqnarray}
J=
\begin{pmatrix}
{\bm 1}&0\\
0&-{\bm 1}
\end{pmatrix}.
\end{eqnarray}
The second term in Eq.~(\ref{ChPT_Lag}) corresponds to the explicit chiral symmetry-breaking term, which provides the finite mass for the NG bosons $\pi^i$.

\subsection{ChPT in the baryon superfluid phase}
\label{sec:ChPTAnalytic}
As demonstrated by the lattice simulations in QC$_2$D, the diquark baryonic matter is created, accompanied by the phase transition to the baryon superfluid phase, at which the chemical potential takes the half value of the vacuum pion mass, $\mu_q^{\rm cr}\equiv m_\pi/2$~\cite{Hands:2001ee,Braguta:2016cpw,Iida:2019rah}. These notable properties can also be described within the chiral effective models including the ChPT.
In this subsection, 
we recapitulate the well-known analytic expressions of the
thermodynamic quantities in the ChPT~\cite{Son:2000by,Hands:2006ve}: the pressure, the quark-number density, the quark-number susceptibility and the energy density.

To evaluate the superfluid phase transition in the ChPT, we take the vacuum expectation value (VEV) for the chiral field $U$,
\begin{eqnarray}
\langle U\rangle
=e^{i2\sqrt{2}X_5 \beta} E^T
=\left( \cos\beta + i2\sqrt{2}X_5 \sin\beta    \right) E^T, \label{UVEV}
\end{eqnarray}
where $\beta$ lies in a range of $0\leq \beta\leq\pi/2$.
By considering the matching with the underlying QC$_2$D, we have the following correspondence between the quark bilinear condensates and $\beta$ in the VEV of $U$~\cite{Kogut:1999iv,Kogut:2000ek}, 
\begin{eqnarray}
\bar{\phi}_{\rm ChPT} = 
\cos \beta,\;\;\;
\bar{d}_{\rm ChPT}=
\sin \beta, \label{CosSinBeta}
\end{eqnarray}
where $\bar{\phi}$ and $\bar{d}$ are defined such that they are normalized by the vacuum chiral condensate $\langle\bar{\psi}\psi\rangle|_{\rm vac}$ as
\begin{eqnarray}
&& \bar{\phi} \equiv \langle \bar \psi \psi\rangle/\langle\bar{\psi}\psi\rangle|_{\rm vac}, \nonumber\\
&& \bar{d} \equiv \langle-\frac{i}{2}\psi^T C\gamma_5 \tau_c^2\tau_f^2 \psi + {\rm H.c.}
\rangle/\langle\bar{\psi}\psi\rangle|_{\rm vac} . \label{ChiDeltaChPT}
\end{eqnarray}
In Eq.~(\ref{CosSinBeta}), the subscript ``ChPT'' has been attached to emphasize that this is the ChPT evaluation for later convenience.
Within the ChPT analysis,
the baryon superfluid phase transition is induced by a nonzero value of $\beta$ as seen from Eqs.~(\ref{CosSinBeta}) and~(\ref{ChiDeltaChPT}).

In this study, we adopt the mean-field approximation where the hadronic loop corrections are not considered. Substituting Eq.~(\ref{ChiDeltaChPT}) into the ChPT Lagrangian~(\ref{ChPT_Lag}) at the mean-field level, an effective potential $V^{\rm mean} = -\langle{\cal L}\rangle$ is obtained as
\begin{eqnarray}
V_{\rm ChPT}^{\rm mean}
(\beta)
=-4 \mu_q^2 f_\pi^2 \sin^2\beta 
- 2f_\pi^2m_\pi^2\cos\beta. \nonumber\\
\label{VEffChPT}
\end{eqnarray}
The phase transition is
determined by solving the stationary condition of
$V_{\rm ChPT}^{\rm mean}$ with respect to $\beta$: $\partial V_{\rm ChPT}^{\rm mean}/\partial \beta =0$, which yields
\begin{eqnarray}
\beta &=&0\;\;\;
\mbox{(for $\mu_q < \mu_q^{\rm cr}$)}
, \nonumber\\
\cos\beta&=&\frac{m_\pi^2}{4 \mu_q^2 }\;\;\;
\mbox{(for $\mu_q\geq \mu_q^{\rm cr}$)}.
\label{cons_ChPT}
\end{eqnarray}
This expression surely shows that the baryon superfluid phase transition in the ChPT occurs at $\mu_q^{\rm cr}=m_\pi/2$. In terms of the chiral and diquark condensates, Eq.~(\ref{cons_ChPT}) implies
\begin{eqnarray}
&& \bar{\phi}^{(\rm H)}_{\rm ChPT} = 1 \ , \ \  \bar{d}^{(\rm H)}_{\rm ChPT} = 0 ,
 \nonumber\\
&& \bar{\phi}^{(\rm BS)}_{\rm ChPT} = \frac{1}{\bar{\mu}^2}\ , \ \  \bar{d}^{(\rm BS)}_{\rm ChPT} = \sqrt{1-\frac{1}{\bar{\mu}^4}}  , 
  \label{NormCondensatesChPT}
\end{eqnarray}
in the hadronic (H) and baryon superfluid (BS) phases, with 
$\bar \mu$ denoting the chemical potential normalized by the critical value,
$\bar \mu = \mu_q/\mu_q^{\rm cr}$. We note that the sign of the diquark condensate has been chosen to be positive. Equation~(\ref{NormCondensatesChPT}) indicates that the chiral condensate is always constant in the hadronic phase. In the superfluid phase, it decreases proportionally to $1/\mu_q^2$. Meanwhile, the diquark condensate approaches $\langle\bar{\psi}\psi\rangle|_{\rm vac}$ as $\mu_q\to\infty$.

By inserting the solution of $\beta$ into the effective potential~(\ref{VEffChPT}),
one can express the pressure, $p= -V^{\rm mean}$, in both the phases:
\begin{eqnarray}
p_{\rm ChPT}^{(\rm H)}&=&
2f_\pi^2 m_\pi^2 ,
\nonumber\\
p_{\rm ChPT}^{(\rm BS)}&=&
f_\pi^2 m_\pi^2\left (
\bar \mu^2+ \frac{1}{\bar \mu^2}
\right) .
\label{PressureChPT}
\end{eqnarray}
In the hadronic phase, the pressure $p_{\rm ChPT}^{(\rm H)}$ does not have the chemical potential dependence and keeps the vacuum value, ensuring that 
the ChPT result satisfies the so-called Silver-Braze property observed in the lattice simulations. At asymptotically large $\mu_q$, $p_{\rm ChPT}^{\rm (BS)}$ grows with a power of $\mu_q^2$.

At the vacuum, the pressure is not reduced to zero but has a finite contribution as in Eq.~(\ref{PressureChPT}), which should be subtracted to facilitate the following demonstration appropriately. After this subtraction, the pressure in the superfluid phase reads
\begin{eqnarray}
p_{\rm ChPT}^{\rm sub} &=& 
p_{\rm ChPT}^{(\rm BS)} -p_{\rm ChPT}^{(\rm H)}
\nonumber\\
&=&
f_\pi^2 m_\pi^2
\left(
\bar \mu -\frac{1}{\bar{\mu}}
\right)^2 .
\label{SubPressureChPT}
\end{eqnarray}
Using this subtracted pressure, 
the baryon number density $n$ and the baryon susceptibility $\chi$ are analytically evaluated as
\begin{eqnarray}
n_{\rm ChPT}&=&
\frac{\partial p_{\rm ChPT}^{\rm sub}}{\partial \mu_q}
=
\frac{1}{\mu_q}2
f_\pi^2 m_\pi^2
\left (
\bar \mu^2
-\frac{1}{\bar \mu^2}
\right),\nonumber\\
\chi_{\rm ChPT}&=& 
\frac{\partial^2p_{\rm ChPT}^{\rm sub}}{\partial \mu_q^2}
=
8f_\pi^2 \left( 1  +   3\frac{1}{\bar \mu^4 }  \right) . 
\label{density_sus_ChPT}
\end{eqnarray}
Furthermore,
the subtracted energy density, $\epsilon = -p +\mu_q n$,  can also be evaluated as
 \begin{eqnarray}
\epsilon_{\rm ChPT}^{\rm sub}
&=&
-p_{\rm ChPT}^{\rm sub} + \mu_q 
n_{\rm ChPT}
\nonumber\\
&=&f_\pi^2 m_\pi^2
\left[
\frac{1}{\bar\mu^2}
(\bar \mu^2+3)
(\bar \mu^2-1)
\right].
\end{eqnarray}

\subsection{
Sound velocity in ChPT}

Using the thermodynamic quantities shown in Sec.~\ref{sec:ChPTAnalytic}, 
one can obtain an analytic expression of the sound velocity in the ChPT within the mean-field approximation.
In general, the sound velocity is defined by 
a derivative of the pressure $p$ with respect to the energy density $\epsilon$:
$c_s^2 = \partial p/\partial \epsilon$, along the isentropic curve. At zero temperature, the isentropic trajectory sticks to the $\mu_q$ axis, and then, the sound velocity can be simplified to the following concise expression~\cite{Minamikawa:2020jfj}: 
\begin{eqnarray}
c_s^2 =\frac{n}{\mu_q \chi}.
\end{eqnarray}

Using the ChPT evaluations of the baryon number density and the baryon susceptibility in Eq.~(\ref{density_sus_ChPT}), the sound velocity is expressed as~\cite{Son:2000by,Hands:2006ve} 
\begin{eqnarray}
\left(c_s^{\rm ChPT} \right)^2 =  
\frac{n_{\rm ChPT} }{ \mu_q \chi_{\rm ChPT} } =
\frac{1 - 
1/\bar \mu^4
}{1 
+3/\bar \mu^4
}
. \label{CsChPT}
\end{eqnarray}
It is interesting to note that
$\left(c_s^{\rm ChPT} \right)^2$ is simply described by the chemical potential tagged with the pion mass (or viewed as the function of only the normalized chemical potential $\bar \mu = \mu_q/\mu_q^{\rm cr}=2\mu_q/m_\pi$).
In the baryonic matter, the sound velocity starts from zero at $\mu_q = \mu_q^{\rm cr}$ (corresponding to $\bar \mu=1$). 
As the chemical potential increases, the sound velocity in the ChPT grows monotonically and approaches $1$ for $\mu_q\to\infty$ without exhibiting a peak.

The $\mu_q$ dependence of the sound velocity has been observed in the QC$_2$D lattice simulation~\cite{Iida:2022hyy}, indicating that the sound velocity departs from zero in the superfluid phase and exceeds the conformal limit $c_s^2=1/3$. 
This conformal limit must be realized for sufficiently large $\mu_q$, since in such a dense matter the relevant scale is solely provided by $\mu_q$. Hence, the lattice result claims the appearance of the peak structure in the intermediate chemical potential regime in cold dense QC$_2$D. This observation, however, contradicts the ChPT estimation~(\ref{CsChPT}). In what follows, we show that this lack of the peak structure in the ChPT framework can be resolved by including contributions from chiral partners of the NG bosons.

\section{LSM in cold dense QC$_2$D}
\label{sec:LSMFramework}

In Sec.~\ref{sec:ChPTReview}, we have shown that the sound velocity evaluated within the ChPT fails in yielding the peak structure. Since the ChPT framework describes only NG boson dynamics, one may infer that the shortcomings are due to the lack of higher excitations. Here, the so-called chiral partners of the NG bosons carrying opposite parities can be considered reasonably as the excitations, which are connected to the NG bosons under the chiral transformation.\footnote{In this regard, the ChPT would be regarded as the ``lowest-energy'' effective model of QC$_2$D where the chiral partners of the NG bosons are integrated out, owing to the nonlinear representation. } To take into account the contribution from these chiral partners, here,
we move onto the LSM approach.


\subsection{LSM at mean-field approximation}
\label{sec:LSMMeanField}

The chiral partners can be introduced by linearizing the chiral field $U=\xi E^T\xi^T = \xi^2 E^T$ in the ChPT, i.e., by parametrizing $\xi^2$ with positive-parity mesons (negative-parity diquark baryons) as well as the NG bosons. Hence, a building block in the LSM incorporating those hadrons can be introduced as~\cite{Suenaga:2022uqn}  
\begin{eqnarray}
\Sigma= (\mathscr{S}^i -i\mathscr{P}^i)X^i E , \label{SIgmaLSM}
\end{eqnarray}
where $X^{i=0}={\bm 1}_{4\times 4}/(2\sqrt{2})$, and $X^i$ ($i=1$ - $5$) is defined in Eq.~(\ref{GeneratorsX}). 
The dynamical variables $\mathscr{S}^i$ and $\mathscr{P}^i$ are identified to the hadrons as follows: $\mathscr{S}^{i=0}$ is the isosinglet scalar meson ($\sigma$ meson);
$\mathscr{S}^{i =1,2,3}$ are the isotriplet scalar mesons ($a_0$ mesons);
$\mathscr{S}^{i=4,5}$ are the positive-parity (anti)baryons;
$\mathscr{P}^{i =0}$ is the isosinglet pseudoscalar meson ($\eta$ meson);
$\mathscr{P}^{i =1,2,3}$ are the isotriplet pseudoscalar meson (pions); 
$\mathscr{P}^{i=4,5}$ are the negative-parity (anti)baryons. It should be noted that $\Sigma$ has mass dimension $+1$ while $U$ is dimensionless, where these fields are related by $\Sigma \sim f_\pi U = f_\pi\xi^2 E^T$.

Given that the difference between $\Sigma$ and $U$ is just the way of parametrization of the hadron field,
$\Sigma$ also transforms under the Pauli-G\"ursey $SU(4)$ symmetry as
\begin{eqnarray}
\Sigma \to g \Sigma g^T,
\end{eqnarray}
following the transformation law of $U$. Hence, with the linearly parametrized $\Sigma$,
the LSM Lagrangian, that is invariant under $SU(4)$ transformation, is constructed as~\cite{Suenaga:2022uqn}
\begin{eqnarray}
{\cal L}_{\rm LSM}
&=&
{\rm tr }[D_\mu\Sigma^\dagger D^\mu \Sigma ]
-m_0^2 {\rm tr}[\Sigma^\dagger \Sigma]
-\lambda_1({\rm tr}[\Sigma^\dagger \Sigma])^2 \nonumber\\
&&
-\lambda_2{\rm tr}[(\Sigma^\dagger \Sigma)^2] +\frac{m_l\bar c}{2}\,{\rm tr}[ E^\dagger \Sigma +\Sigma^\dagger E ]
\ ,
\label{LSM_lag}
\end{eqnarray}
where 
$m_0^2$ is a mass parameter, and $\lambda_1$ and $\lambda_2$ are dimensionless parameters controlling interactions among the hadrons. The last piece proportional to $m_l\bar{c}$ is responsible for the explicit breaking of the Pauli-G\"ursey $SU(4)$ symmetry which is inevitable to generate finite masses of the NG bosons. In Eq.~(\ref{LSM_lag}), the $U(1)_A$ anomalous contributions are not introduced since the lattice results on the mass difference between $\eta$ and pion imply small effects from the $U(1)_A$ anomaly~\cite{Suenaga:2022uqn}.

According to the matching with the underlying QC$_2$D, the fields $\mathscr{S}^0$ and $\mathscr{P}^5$ can be linked to the quark bilinear fields~\cite{Kawaguchi:2023olk}:
\begin{eqnarray}
\bar \psi \psi =  -\sqrt{2}\bar{c}\mathscr{S}^0, \;\;\; -\frac{i}{2}
\psi^T C\gamma_5 \tau_c^2\tau_f^2 \psi + {\rm H.c.}
=-\sqrt{2}\bar{c} \mathscr{P}^5. \nonumber\\
\label{MatchingLSM}
\end{eqnarray}
Within the LSM, 
the chiral condensate and the diquark condensate are, hence, introduced as
 \begin{eqnarray}
\sigma_0
\equiv
\langle \mathscr{S}^0 \rangle\ , \ \ \Delta\equiv \langle \mathscr{P}^5 \rangle\ ,
\label{delcon} 
\end{eqnarray}
where 
$\sigma_0$ and $\Delta$ are chosen to be positive.
As in the case of the ChPT analysis,
we also implement the mean-field approximation in the present LSM analysis. 
Then, substituting Eq.~(\ref{delcon}) into Eq.~(\ref{LSM_lag}) through Eq.~(\ref{SIgmaLSM}), the effective potential is expressed as
\begin{eqnarray}
V^{\rm mean}_{\rm LSM}(\sigma_0,\Delta) &=&
-2\mu_q^2\Delta^2+\frac{m_0^2}{2}(\sigma_0^2+\Delta^2) \nonumber\\
&& 
+\frac{\tilde{\lambda}}{4}(\sigma_0^2+\Delta^2)^2
-\sqrt{2}m_l\bar c\sigma_0\ ,
\label{effpot_LSM}
\end{eqnarray}
where we have defined $\tilde{\lambda} \equiv \lambda_1+\lambda_2/4$ to denote the four-point couplings collectively.
The phase structure can be evaluated through stationary conditions,
\begin{eqnarray}
\frac{V^{\rm mean}_{\rm LSM}(\sigma_0,\Delta)}{\partial \sigma_0}=0,\;\;\;
\frac{V^{\rm mean}_{\rm LSM}(\sigma_0,\Delta)}{\partial \Delta}=0 .
\label{st_con_LSM}
\end{eqnarray}
By solving these stationary conditions, one can see that the critical chemical potential to enter the superfluid phase coincides with the one derived in the ChPT: $\mu_q^{\rm cr} = m_\pi/2$. In fact, from Eq.~(\ref{st_con_LSM}) the condensates are evaluated as 
\begin{eqnarray}
&&\sigma_0^{(\rm H)} = \sqrt{2}m_l\bar c/m_\pi^2 , \nonumber\\
&&\Delta^{(\rm H)}=0 \ \ \ \ \ \ \ \ \mbox{(for $\mu_q< \mu_q^{\rm cr}$)} , 
\end{eqnarray}
while
\begin{eqnarray}
&&\sigma_0^{(\rm BS)} 
 =\frac{\sigma_{0}^{\rm (H)}
 m_{\pi}^2
 }{4\mu_q^2} , \nonumber\\
 && \Delta^{(\rm BS)} =
\left[
\frac{ 1}{\tilde{\lambda}}
(4 \mu_q^2 -m_0^2)-(\sigma_0^{(\rm BS)})^2\right]^{1/2}
\;\;\;
\mbox{(for $\mu_q\geq \mu_q^{\rm cr}$)} . \nonumber\\
\label{cons_LSM}
\end{eqnarray}
We note that the vacuum pion mass within the LSM is read from the quadratic term of the pion
fluctuation field upon the mean-field approximation, in Eq.~(\ref{LSM_lag}), as
\begin{eqnarray}
    m_\pi^2 &=& 
m_0^2 + \tilde{\lambda}(\sigma_0^{\rm (H)})^2  .
\label{MPiVac}
\end{eqnarray}
Following a similar procedure, the vacuum mass of the chiral partner of the pion, the $\sigma$ meson, is evaluated to be
\begin{eqnarray}
m_\sigma^2 = m_0^2 + 3\tilde{\lambda}(\sigma_0^{\rm (H)})^2 . \label{MSigmaVac}
\end{eqnarray}

The analytic expressions of $\sigma_0$ and $\Delta$ in Eq.~(\ref{cons_LSM}) include explicit model parameters, which prevent us from achieving an intuitive picture of their $\mu_q$ dependencies. For this reason, in what follows we try to translate $\sigma_0^{\rm (H)}$, $m_0^2$ and $\tilde{\lambda}$ into more physically transparent quantities. First, by evaluating the broken current associated with the broken $SU(4)/Sp(4)$ space, one can easily show that the mean field $\sigma_0^{\rm (H)}$ is connected to the pion decay constant as
\begin{eqnarray}
\sigma_0^{\rm (H)} = \sqrt{2} f_\pi, \label{SigmaFPi}
\label{sigma_decay_con}
\end{eqnarray}
which is derived in the Appendix. 
Next, from Eqs.~(\ref{MPiVac}) and~(\ref{MSigmaVac}), $m_0^2$ and $\tilde{\lambda}$ can be expressed in terms of the pion and $\sigma$ meson masses, which reads
\begin{eqnarray}
m_0^2 = \frac{3m_\pi^2-m_\sigma^2}{2} \ , \ \ \tilde{\lambda} = \frac{m_\sigma^2-m_\pi^2}{2(\sigma_0^{\rm (H)})^2} .
\end{eqnarray}
Therefore, substituting those relations into Eq.~(\ref{cons_LSM}), finally $\sigma_0$ and $\Delta$ in the superfluid phase are rewritten to
\begin{eqnarray}
&&\sigma_0^{(\rm BS)} = \frac{\sqrt{2}f_\pi}{\bar{\mu}^2} , \nonumber\\
&&\Delta^{(\rm BS)} = \left[\frac{16f_\pi^2}{\delta \bar{m}_{\sigma-\pi}^2}(\bar{\mu}^2-1) + 2f_\pi^2\left(1-\frac{1}{\bar{\mu}^4}\right) \right]^{1/2} , \nonumber\\
\label{MeanFieldPhysical}
\end{eqnarray}
where the normalized chemical potential $\bar{\mu} = \mu_q/\mu_q^{\rm cr}$ is used, and the dimensionless mass difference between $\sigma$ meson and pion is defined by\footnote{A mass difference of the chiral partners, the $\sigma$ meson and pion, measures strength of the chiral symmetry breaking. When the chiral symmetry is explicitly and/or spontaneously broken, the $\sigma$ meson mass deviates from the pion mass, leading to the nonzero value of $\delta \bar m^2_{\sigma-\pi}$.
}
\begin{eqnarray}
\delta \bar{m}_{\sigma-\pi}^2 \equiv \frac{m_\sigma^2-m_\pi^2}{(\mu_q^{\rm cr})^2} . \label{DeltaMDef}
\end{eqnarray}
Equation~(\ref{MeanFieldPhysical}) clearly shows that the fate of the mean fields $\sigma_0$ and $\Delta$ in the superfluid phase is essentially determined by vacuum values of the decay constant and the mass difference between the chiral partners.\footnote{This simple expression of the mean fields is not modified even when the $U(1)_A$ anomaly effects are introduced, and accordingly the following discussions are not changed.}

At the end of this subsection, we make comparisons between the ChPT and LSM results on $\mu_q$ dependencies of the chiral and diquark condensates. From the matching result in Eq.~(\ref{MatchingLSM}) together with Eq.~(\ref{MeanFieldPhysical}), one can derive the following relations with respect to the normalized chiral and diquark condensates in the superfluid phase:
\begin{eqnarray}
&& \bar{\phi}^{(\rm BS)}_{\rm LSM} = \bar{\phi}^{(\rm BS)}_{\rm ChPT}, \nonumber\\
&&( \bar{d}^{(\rm BS)}_{\rm LSM})^2 = (\bar{d}^{(\rm BS)}_{\rm ChPT})^2 + (\delta \bar{d}^{\rm (BS)})^2 , \label{CondensatesComparison}
\end{eqnarray}
with
\begin{eqnarray}
( \delta\bar{d}^{(\rm BS)})^2 \equiv \frac{8}{\delta\bar{m}_{\sigma-\pi}^2}(\bar{\mu}^2-1) , \label{AdditionalD}
\end{eqnarray}
where the ChPT results on $\bar{\phi}$ and $\bar{d}$ are provided in Eq.~(\ref{NormCondensatesChPT}). Thus, from Eq.~(\ref{CondensatesComparison}), the normalized chiral condensate is found to share the common $\mu_q$ scaling in the superfluid phase. Meanwhile, the squared diquark condensate is divided into the ChPT result and an additional contribution $( \delta\bar{d}^{(\rm BS)})^2$. In particular, Eq.~(\ref{AdditionalD}) indicates that the additional effects are quantified by the mass difference of the chiral partners, $\sigma$ meson and pion, and amplified as the chemical potential increases. Since $( \delta\bar{d}^{(\rm BS)})^2$ is proportional to $1/\delta\bar{m}_{\sigma-\pi}^2$, $\bar{d}^{(\rm BS)}_{\rm LSM}$ is reduced to $\bar{d}^{(\rm BS)}_{\rm ChPT}$ when the $\sigma$ meson mass is sufficiently large compared to the pion mass. This reduction is natural; such a heavy-mass limit is equivalent to integrating out the $\sigma$ meson from the LSM framework, which should converge on the ChPT. Hence, one can conclude that the $( \delta\bar{d}^{(\rm BS)})^2$ is induced as a correction beyond the lowest-energy dynamics governed by the NG bosons. Those clear properties of the LSM played by the $\sigma$ meson are also reflected to the thermodynamic quantities, as is shown in the following analysis.

\subsection{Thermodynamic quantities in LSM}
\label{sec:ThermodynamicsLSM}

In this subsection, we exhibit the thermodynamic quantities in the LSM approach within the mean-field approximation, and shed light on connections between results from the ChPT and LSM.

The pressure within the LSM is straightforwardly evaluated from the potential~(\ref{effpot_LSM}). In terms of the physical quantities, $m_\pi$, $m_\sigma$ and $f_\pi$ provided in Eqs.~(\ref{MPiVac}),~(\ref{MSigmaVac}), and~(\ref{SigmaFPi}), it reads 
\begin{eqnarray}
p_{\rm LSM}^{(\rm H)} 
&=&
f_\pi^2 m_\pi^2
\left(\frac{\delta \bar m^2_{\sigma-\pi}}{16}+1\right) ,\nonumber\\
p_{\rm LSM}^{(\rm BS)} 
&=& f_\pi^2 m_\pi^2
\Bigg[
\left(\frac{4}{\delta \bar m^2_{\sigma-\pi}}+\frac{1  }{\bar \mu^2}\right)(\bar \mu^2 -1 )^2 \nonumber\\
&& 
+\left(\frac{\delta \bar m^2_{\sigma-\pi}}{16}+1\right)
\Bigg] ,
\label{pressure_LSM}
\end{eqnarray}
in both the hadronic and baryon superfluid phases. Similarly to the normalized diquark condensate indicated in Eq.~(\ref{CondensatesComparison}), the pressure in both the phases includes the mass difference of the chiral partners $\delta \bar m^2_{\sigma-\pi}$. We note that the LSM analysis surely satisfies the Silver-Braze property in the hadronic phase, as seen from Eq.~(\ref{pressure_LSM}).



Here, we discuss connections between the LSM and ChPT results in terms of the thermodynamic quantities. 
Subtracting the vacuum pressure from $p_{\rm LSM}^{(\rm BS)}$ similarly to Eq.~(\ref{SubPressureChPT}), one can find that the subtracted pressure evaluated in the LSM is also separated into the ChPT result and an additional part:
\begin{eqnarray}
p_{\rm LSM}^{\rm sub}
&=&
p_{\rm LSM}^{\rm (BS)} - p_{\rm LSM}^{\rm (H)}
\nonumber\\
&=&
 p_{\rm ChPT}^{\rm sub} 
+\delta  p, \label{SubPressureLSM}
\end{eqnarray}
where $\delta p$ reads
\begin{eqnarray}
\delta p
=
f_\pi^2 m_\pi^2
\left[
\frac{4}{\delta \bar m^2_{\sigma-\pi}}(\bar \mu^2 -1 )^2
\right].   
\end{eqnarray}
In this expression a square bracket has been included to extract the common coefficient derived in the ChPT result~(\ref{SubPressureChPT}).
Accordingly, the baryon number density and the baryon susceptibility in the LSM are found to be linked with the ChPT results,
\begin{eqnarray}
 n_{\rm LSM}  &=&
n_{\rm ChPT} 
+
\delta n,
\nonumber\\
\chi_{\rm LSM} &=&
\chi_{\rm ChPT}+
\delta \chi,
\label{density_sus_LSM}
\end{eqnarray}
where $\delta n$ and $\delta \chi$ are given by
\begin{eqnarray}
\delta n&=&
\frac{1}{\mu_q}
2 f_\pi^2  m_\pi^2
\left[
\frac{8}{\delta \bar m^2_{\sigma-\pi}}(\bar \mu^4-\bar \mu^2)
\right],\nonumber\\
\delta \chi
&=&
8f_\pi^2
\left[
\frac{8}{\delta \bar m^2_{\sigma-\pi}} (3\bar \mu^2-1)
\right].
\end{eqnarray}
In addition, 
we can evaluate the subtracted energy in the LSM from $\epsilon=-p+\mu_qn$, which yields 
\begin{eqnarray}
\epsilon_{\rm LSM}^{\rm sub} 
&=&
\epsilon_{\rm ChPT}^{\rm sub} +
\delta\epsilon,
\end{eqnarray}
where $\delta\epsilon$ is given by
\begin{eqnarray}
\delta\epsilon &=& f_\pi^2
m_\pi^2
\left[
\frac{4}{\delta \bar m^2_{\sigma-\pi}}(3\bar \mu^2+1)
(\bar \mu^2 -1)
\right]. \nonumber\\
\end{eqnarray}

The contributions, $\delta p$, $\delta n$, $\delta \chi$ and $\delta \epsilon$, are expressed by the pion decay constant, pion mass and mass difference of the chiral partners, as well as the chemical potential.
Notably, these terms are proportional to the inverse of $\delta \bar m^2_{\sigma-\pi}$:
\begin{eqnarray}
\delta p, \delta n, \delta \chi, \delta \epsilon \propto 1/\delta \bar m^2_{\sigma-\pi}.
\end{eqnarray}
At the heavy mass limit of the $\sigma$ meson, 
those corrections are suppressed such that
the thermodynamic quantities are reduced to the ChPT results:
\begin{eqnarray}
\lim_{m _{\sigma}\to\infty}
p_{\rm LSM}^{\rm sub} &=& p_{\rm ChPT}^{\rm sub},\nonumber\\
\lim_{m _{\sigma}\to\infty}
 n_{\rm LSM}  &=&
n_{\rm ChPT}, 
\nonumber\\
\lim_{m _{\sigma}\to\infty}
\chi_{\rm LSM} &=&
\chi_{\rm ChPT},\nonumber\\
\lim_{m _{\sigma}\to\infty}
\epsilon_{\rm LSM}^{\rm sub} 
&=&
\epsilon_{\rm ChPT}^{\rm sub},
\end{eqnarray}
as observed for the squared diquark condensate in Eq.~(\ref{CondensatesComparison}). Therefore, as explained at the end of Sec.~\ref{sec:LSMMeanField}, the additional contributions for the thermodynamic quantities are regarded as corrections beyond the lowest-energy regime governed by the NG bosons.

\subsection{
Sound velocity in LSM
}
\label{sec:SoundVelocityLSM}

In Sec.~\ref{sec:ThermodynamicsLSM},
we have evaluated the thermodynamic quantities 
by employing the LSM and succeeded in
extending the ChPT results so as to include the corrections from the chiral partners.
Here, we discuss the chiral partner contribution to the sound velocity within the LSM framework.

Using the baryon number density and the baryon susceptibility in Eq.~(\ref{density_sus_LSM}), 
we find the analytical expression of the sound velocity in the LSM within the mean-field approximation,
\begin{eqnarray}
\left(c_s^{\rm LSM} \right)^2 &=&  
\frac{n_{\rm ChPT} +\delta n}{ \mu_q (\chi_{\rm ChPT} +\delta \chi)}
\nonumber\\
&=&
\frac{ (1-1/\bar\mu^4) +8(\bar \mu^2-1)/\delta \bar m^2_{\sigma-\pi} }{(1+3/\bar\mu^4)+8(3\bar \mu^2-1)/\delta \bar m^2_{\sigma-\pi}   },
\label{sound_LSM}
\end{eqnarray}
which is only dependent on the masses of $\sigma$ meson and pion together with the chemical potential but is independent of the pion decay constant.

When we take $m_\sigma\to\infty$ while keeping $\mu_q$ finite, the sound velocity is reduced to the ChPT result:
\begin{eqnarray}
\lim_{m_\sigma\to \infty}\left(c_s^{\rm LSM}\right)^2 = \left(c_s^{\rm ChPT}\right)^2 = \frac{1-1/\bar{\mu}^4}{1+3/\bar{\mu}^4} ,
\end{eqnarray}
as naively expected. On the other hand, when taking $\mu_q\to\infty$ with $m_\sigma$ kept finite, the sound velocity approaches the conformal limit in the end:
\begin{eqnarray}
\lim_{\mu_q\to \infty}\left(c_s^{\rm LSM}\right)^2  = \frac{1}{3}. \label{CsLimitConformal}
\end{eqnarray}
We note that the smaller value of $m_\sigma$ taken the more rapidly the sound velocity converges on the conformal limit.


Although $\left(c_s^{\rm LSM}\right)^2$ includes the $\delta\bar{m}_{\sigma-\pi}^2$ contributions nonlinearly, formally one can decompose $\left(c_s^{\rm LSM}\right)^2$ into the ChPT result and corrections as 
\begin{eqnarray}
\left(c_s^{\rm LSM}\right)^2= \left(c_s^{\rm ChPT}\right)^2 + \delta c_s^2 ,
\end{eqnarray}
by explicitly extracting Eq.~(\ref{CsChPT}). After this decomposition, $\delta c_s^2<0$ for any $\bar{\mu}$ $(\geq1)$ is easily proven analytically while, of course, $\left(c_s^{\rm ChPT}\right)^2>0$. Hence, there appears a competition between $\left(c_s^{\rm ChPT}\right)^2>0$ and $\delta c_s^2<0$, i.e., between the ChPT contributions from the NG bosons and the correction provided from the chiral-partner dynamics. As a consequence, the emergence of nonmonotonic behaviors in the sound velocity, including the peak structure, is expected.



\section{Numerical demonstrations}
\label{sec:SoundVelocityLSM}

In Sec.~\ref{sec:LSMFramework}, we have learned that the LSM analysis  
leads to generating corrections to the ChPT results of the thermodynamics quantities, that is characterized by the mass difference $\delta\bar{m}_{\sigma-\pi}^2$. In this section, focusing on this clear structure, we present numerical results on the sound velocity and the trace anomaly to take a closer look at roles of the $\sigma$ meson. We note that currently the mass of the $\sigma$ meson defined as the chiral partner remains obscure, due to not only errors in the lattice results~\cite{Murakami:2022lmq} but also possible mixing effects from other hadrons such as a $0^+$ glueball. Hence, in what follows we vary the value of $m_\sigma$ for numerical demonstrations.

\subsection{Sound velocity}
\label{sec:CsNumerical}

In this subsection, we present numerical results on the impact of the $\sigma$ meson on the sound velocity.

\begin{figure}[H] 
\begin{center}
        \includegraphics[width=8.5cm]{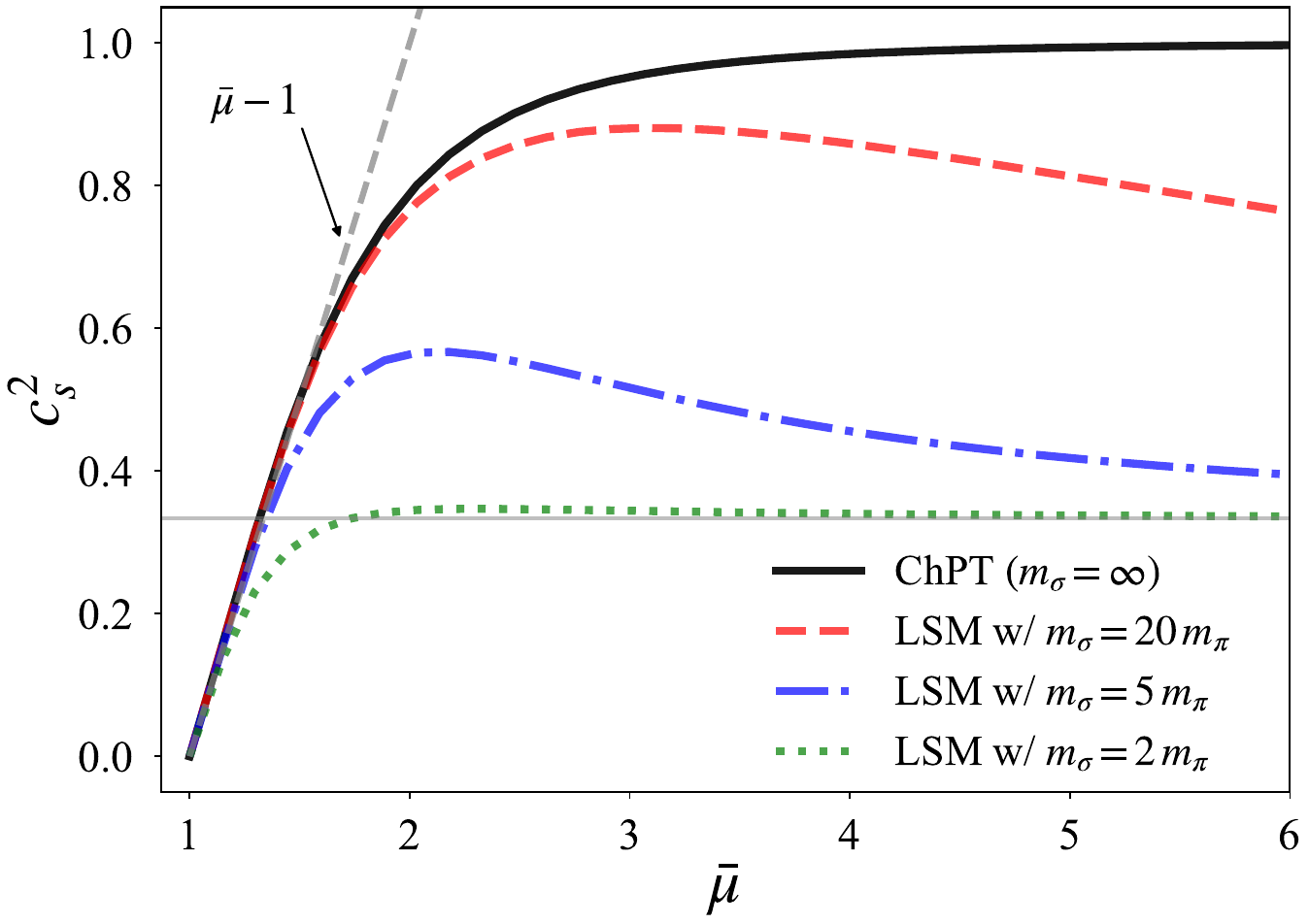}
\end{center}
\caption{$\mu_q$ dependence of the sound velocity with various values of $m_\sigma$.}
\label{sound_LSM_ChPT}
\end{figure}

Depicted in Fig.~\ref{sound_LSM_ChPT} is the resultant $\mu_q$ dependencies of the sound velocities evaluated in the LSM with several values of $m_\sigma$. The solid black curve corresponds to the ChPT result which is also reproduced within the LSM framework by taking $m_\sigma\to\infty$. The peak structure for certain choices of $m_\sigma$ is indeed observed, as the recent lattice data implies~\cite{Iida:2022hyy}. The figure also indicate that, in the vicinity of $\mu_q=\mu_q^{\rm cr}$, the results are less sensitive to the value of $m_\sigma$. In fact, from Eq.~(\ref{sound_LSM}), the sound velocity for $\mu_q\approx\mu_q^{\rm cr}$ is simply approximated by 
\begin{eqnarray}
\left(c_s^{\rm LSM}\right)^2 \approx \bar{\mu}-1 , \label{CsAround1}
\end{eqnarray}
which does not exhibit any $\delta\bar{m}_{\sigma-\pi}^2$ dependencies. This linear line is shown by the dashed line in Fig.~\ref{sound_LSM_ChPT}. On the other hand, when looking at the high-density region such as $\mu_q =3m_\pi \gg \mu_q^{\rm cr}$,
the value of $c_s^2$ changes from $c_s^2=1$ toward
$c_s^2=1/3$ in association with reduction of the $\sigma$ meson mass, indicating a significant dependence on $m_\sigma$. In fact, for finite $m_\sigma$, one can show that the sound velocity in high dense regime is expanded as 
\begin{eqnarray}
\left(c_s^{\rm LSM}\right)^2 = \frac{1}{3} + \frac{\delta\bar{m}_{\sigma-\pi}^2-8}{36}\frac{1}{\bar{\mu}^2} + {\cal O}\left(1/\bar{\mu}^3\right) , \label{CsLSMDense}
\end{eqnarray}
where the coefficient of $1/\bar{\mu}^2$ depends on $m_\sigma$. 
This clearly shows that the sound velocity for $\bar{\mu}\gg1$ is strongly affected by the value of $m_\sigma$, and finally it converges on $1/3$ as long as $m_\sigma$ is not infinitely large, as derived in Eq.~(\ref{CsLimitConformal}).

\begin{figure}[H] 
\begin{center}
        \includegraphics[width=8.5cm]{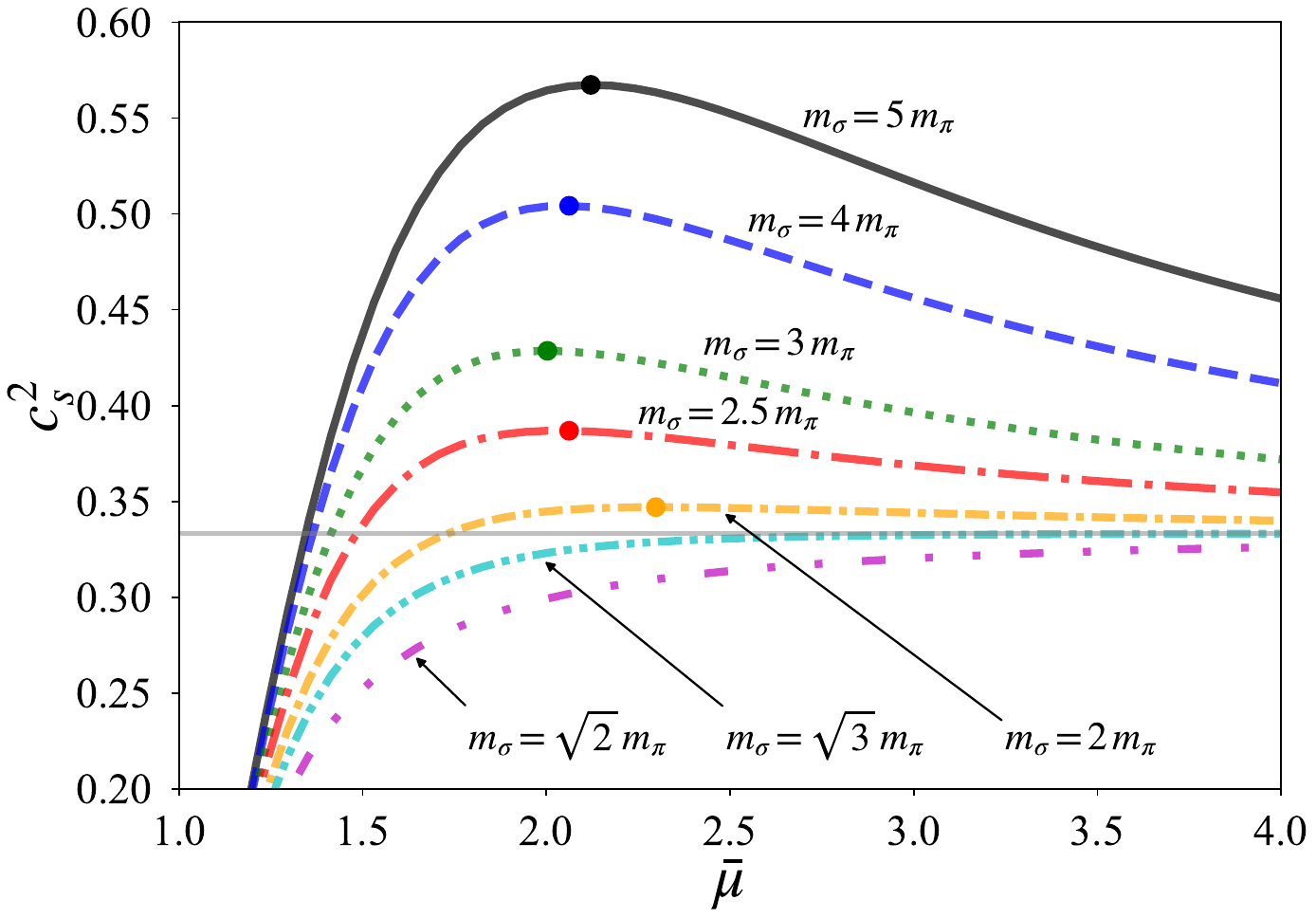}
\end{center}
\caption{ Constraints on the emergence of the sound velocity peak.
}
\label{sound_peak}
\end{figure}

Equation~(\ref{CsLSMDense}) also implies that when the coefficient of $1/\bar{\mu}^2$ is positive, i.e., $\delta\bar{m}_{\sigma-\pi}^2>8$, the sound velocity approaches $1/3$ from above for sufficiently large $\mu_q$, leading to the at least one peak structure. In order to take a closer look the emergence of the peak, here we consider a stationary condition of $c_s^2$: $\left.\partial c_s^2/\partial \bar \mu\right|_{\bar \mu = \bar \mu_p}
=0$, which is, more concretely, rewritten into
\begin{eqnarray}
(
\delta \bar m^2_{\sigma-\pi}
-2\bar \mu^6_p)
(
\delta \bar m^2_{\sigma-\pi}
-8
)
+18 \delta \bar  m^2_{\sigma-\pi}
\bar \mu^2_p=0 . \label{MuBarpEq}
\end{eqnarray}
This equation possesses a single solution of positive $\bar{\mu}_p^2$ only when $\delta \bar m^2_{\sigma-\pi}
-8>0$ is satisfied. In other words, within the present LSM, we can obtain only one peak in the sound velocity and this peak is derived only when
\begin{eqnarray}
m_\sigma > \sqrt{3}m_\pi . \label{PeakCondition}
\end{eqnarray}
On the contrary, when the $\sigma$ meson mass is constrained by $m_\sigma\leq\sqrt{3}m_\pi$, the sound velocity monotonically grows toward the conformal limit $1/3$ without exhibiting any peaks. We can verify the constraint~(\ref{PeakCondition}) numerically, as illustrated in Fig.~\ref{sound_peak}. In this figure, the peak positions indicated by filled circles surely appear only when~(\ref{PeakCondition}) is satisfied. The figure also implies that two different values of $m_\sigma$ can yield the identical peak position $\bar{\mu}_p$. 
Such a doubling of the peak position
is realized because Eq.~(\ref{MuBarpEq}) is a quadratic equation with respect to $\delta\bar{m}_{\sigma-\pi}^2$.


\begin{figure*}[t] 
\begin{tabular}{cc}
\begin{minipage}{0.5\hsize}
\begin{center}
    \includegraphics[width=8.2cm]{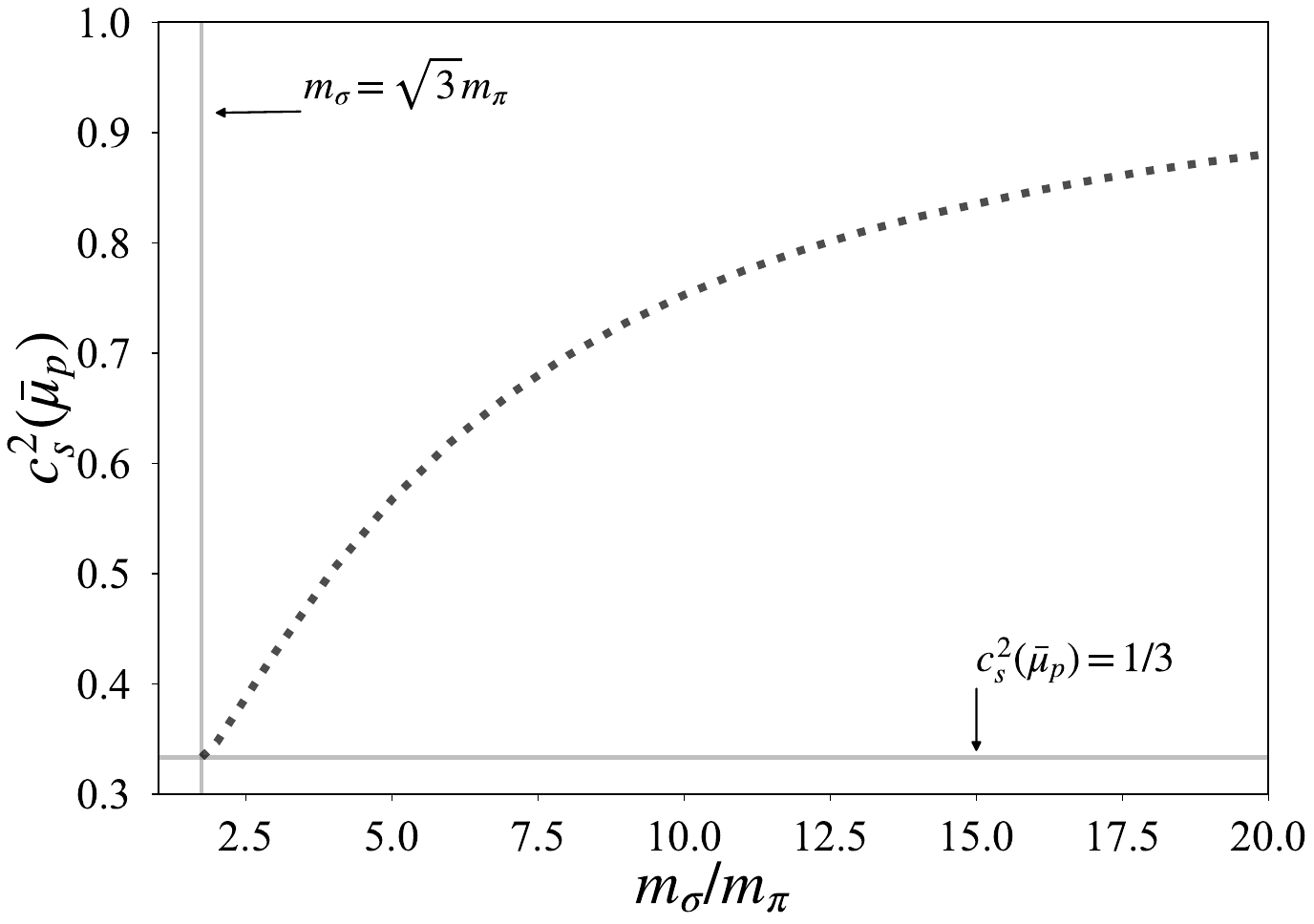}
    \subfigure{(a)}
\end{center}
\end{minipage}
\begin{minipage}{0.5\hsize}
\begin{center}
    \includegraphics[width=8.2cm]{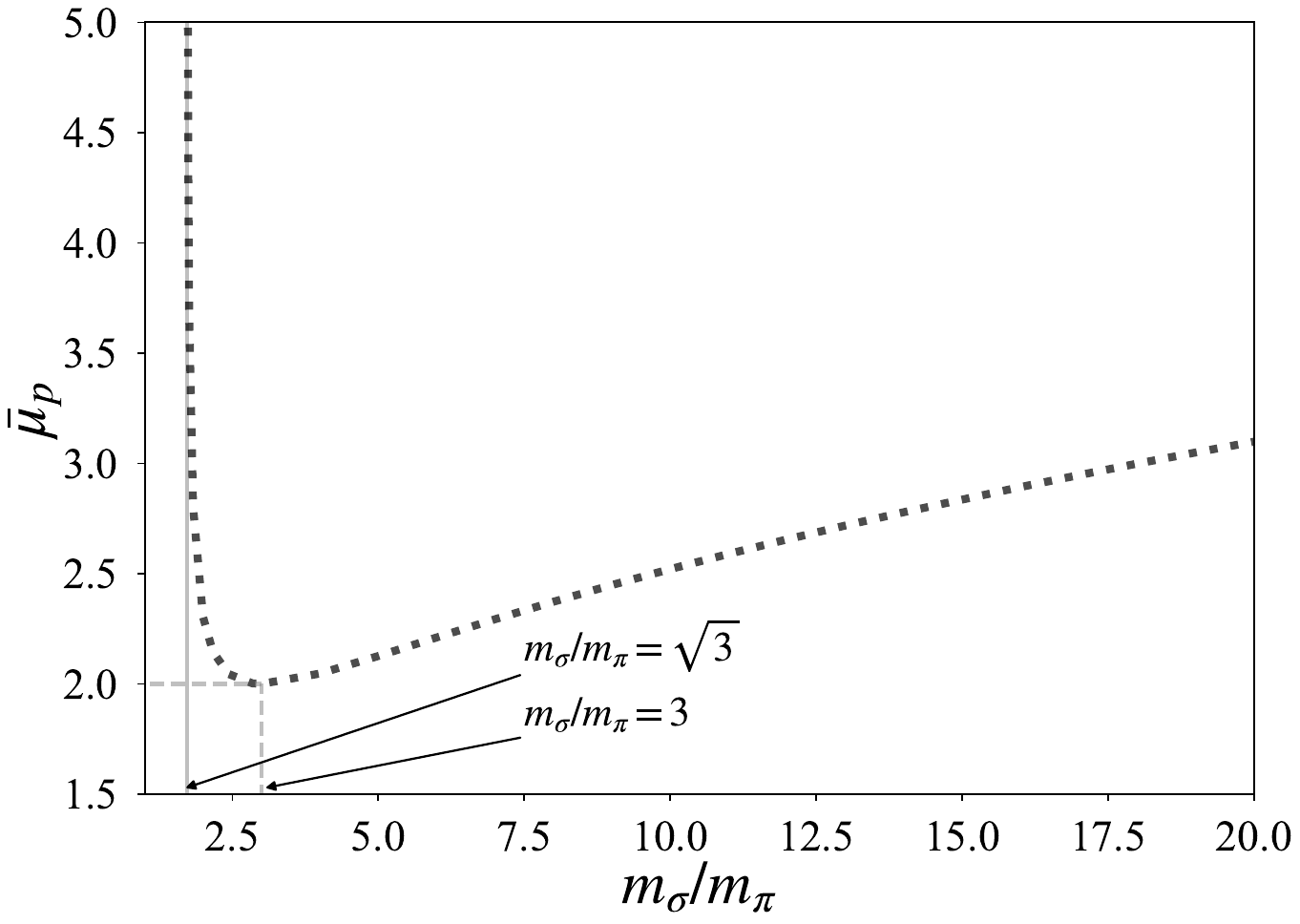}
    \subfigure{(b)}
\end{center}
\end{minipage}
\end{tabular}
\caption{Peak value (a) and peak position (b) as functions of the $\sigma$ meson mass.}
\label{peak}
\end{figure*}

As the $\sigma$ meson mass becomes heavier, the suppression of the sound velocity from the correction $\delta c_s^2$ gets weakened. In order to gain insights into a relationship between this weakening and the peak properties, we depict the peak value $c_s^2(\bar{\mu}_p)$ and the peak position $\bar{\mu}_p$ as functions of $m_\sigma$, in panels (a) and (b) of Fig.~\ref{peak}, respectively. Panel (a) of Fig.~\ref{peak} shows that as $m_\sigma$ becomes larger from the critical value $m_\sigma = \sqrt{3}m_\pi$, the peak value is enhanced from $c_s^2=1/3$ toward $c_s^2=1$ monotonically. From panel (b), one can see that there exists a lower limit of the peak position, $\bar \mu_p =2$, realized when $m_\sigma =3 m_\pi$. In other words, the peak is generated only for $\mu_q > 2\mu_q^{\rm cr}$ within the present LSM approach. Besides, the doubling of the peak position 
explained below Eq.~(\ref{PeakCondition}) is clearly seen in this panel. Despite such a doubling, its peak value is distinct from panel (a) of Fig.~\ref{peak}. Therefore, by combining both the information, the peak value and the peak position, one can identify the value of $m_\sigma$ exclusively. 


\subsection{Trace anomaly}
\label{sec:TraceAnomalyNumerical}

Owing to the contribution of the $\sigma$ meson, the asymptotic behavior of the sound velocity at the high-density region is changed to approach the conformal limit.
The related conformal property would be further investigated from the trace anomaly measured by $\Theta^\mu_\mu=\epsilon-3p$.
In the recent QC$_2$D lattice simulation, the chemical potential dependence of the trace anomaly has also been computed~\cite{Iida:2022hyy}, implying that its sign turns from positive to negative at some chemical potential in the superfluid phase.
In this subsection, motivated by this characteristic property, we explore the correlation between the sound velocity peak and the sign of the trace anomaly in the baryonic matter within the LSM approach.

In the QC$_2$D lattice simulation,
the $\mu_q$ dependence of 
the trace anomaly is evaluated by subtracting the vacuum value from $\Theta^\mu_\mu$~\cite{Iida:2022hyy}. 
Following this lattice definition, we evaluate the subtracted trace anomaly in the LSM as\footnote{Owing to this subtraction, vanishing of the trace anomaly evaluated here, 
$\left(\Theta^{\rm sub}\right)^\mu_\mu=0$, does not correspond to the restoration of the scale symmetry of the theory.
}
\begin{eqnarray}
(\Theta_{\rm LSM}^{\rm sub})^\mu_\mu &=& 
 \epsilon_{\rm LSM}^{\rm sub} -3 p_{\rm LSM}^{\rm sub} \nonumber\\
&=&
(\Theta_{\rm ChPT})^\mu_\mu
+\delta\Theta^\mu_\mu, 
\end{eqnarray}
where the subtracted ChPT contribution $(\Theta_{\rm ChPT}^{\rm sub})^\mu_\mu$ and the chiral-partner correction $\delta\Theta^\mu_\mu$ are provided by
\begin{eqnarray}
(\Theta_{\rm ChPT}^{\rm sub})^\mu_\mu
&=&
\epsilon_{\rm ChPT}^{\rm sub} - 3 p_{\rm ChPT}^{\rm sub}\nonumber\\
&=& 
f_\pi^2 m_\pi^2
\left [
\frac{-2}{\bar \mu^2}(\bar \mu^2-3) (\bar \mu^2-1)
\right], \nonumber\\
\end{eqnarray}
and
\begin{eqnarray}
\delta\Theta^\mu_\mu &=& 
\delta\epsilon -3\delta p  \nonumber\\
&=&
f_\pi^2 m_\pi^2\left[ \frac{16}{\delta \bar m^2_{\sigma-\pi}}  (\bar \mu^2-1)\right],
\end{eqnarray}
respectively. Here, to make sure, we again emphasize that the correction is proportional to $1/\delta\bar{m}_{\sigma-\pi}^2$ which vanishes for $m_\sigma\to\infty$ limit, and hence in this limit $(\Theta_{\rm LSM}^{\rm sub})^\mu_\mu$ is reduced to $(\Theta_{\rm ChPT}^{\rm sub})^\mu_\mu$.
In chiral effective models, the typical mass scale of the trace anomaly is characterized by $f_\pi^2 m_\pi^2$. Then,
to easily grasp the $\mu_q$ dependence of the trace anomaly,
we define the following dimensionless one:
\begin{eqnarray}
\bar \Theta
=\frac{
\Theta^\mu_\mu
}{f_\pi^2 m_\pi^2}.
\end{eqnarray}

Looking at the high-density regions,
the $\mu_q$ scaling of
the normalized trace anomaly $\bar \Theta$ in the LSM reads
\begin{eqnarray}
\bar \Theta_{\rm LSM}
&=&
\left(
\frac{16}{\delta \bar m^2_{\sigma-\pi}} 
-2
\right) \bar \mu^2
+
\left(
-\frac{16}{\delta \bar m^2_{\sigma-\pi}} +
8
\right) \nonumber\\
&& +O(1/\bar\mu^4).  \label{ThetaBar}
\end{eqnarray}
From this expression, we find that in the high-density regions
the sign of trace anomaly is related to the $\sigma$ meson mass. When $m_\sigma=\sqrt{3}m_\pi$ the first term of Eq.~(\ref{ThetaBar}) vanishes, leading to $\bar \Theta_{\rm LSM}\to6$ that is positive. For smaller $\sigma$ meson mass, $m_\sigma<\sqrt{3}m_\pi$, $\bar \Theta_{\rm LSM}$ positively diverges; meanwhile, when the $\sigma$ meson mass satisfies $m_\sigma>\sqrt{3}m_\pi$, the asymptotic sign of $\bar \Theta_{\rm LSM}$ turns out to be negative. It is noteworthy that the critical value for the sign of the trace anomaly $m_\sigma=\sqrt{3}m_\pi$ is identical to the one which discriminates the emergence of the sound velocity peak, as delineated in Sec.~\ref{sec:CsNumerical}. Those remarkable correlations are summarized in Table~\ref{tab:Relation}.

\begin{table}[t]
\begin{center}
  \begin{tabular}{c||c|c}  \hline\hline
$\sigma$ Meson mass & Peak of $c_s^2$ & Asymptotic sign of $\Bar{\Theta}$ \\ \hline
$m_\sigma> \sqrt{3}m_\pi$ & $\checkmark$ & Negative\\ 
$m_\sigma \leq\sqrt{3}m_\pi$ &  & Positive \\
\hline \hline
 \end{tabular}
\caption{Relation between the sound velocity peak and the sign of the trace anomaly for $\bar{\mu}\to\infty$.}
\label{tab:Relation}
\end{center}
\end{table}

\begin{figure}[H] 
\begin{center}
        \includegraphics[width=8.5cm]{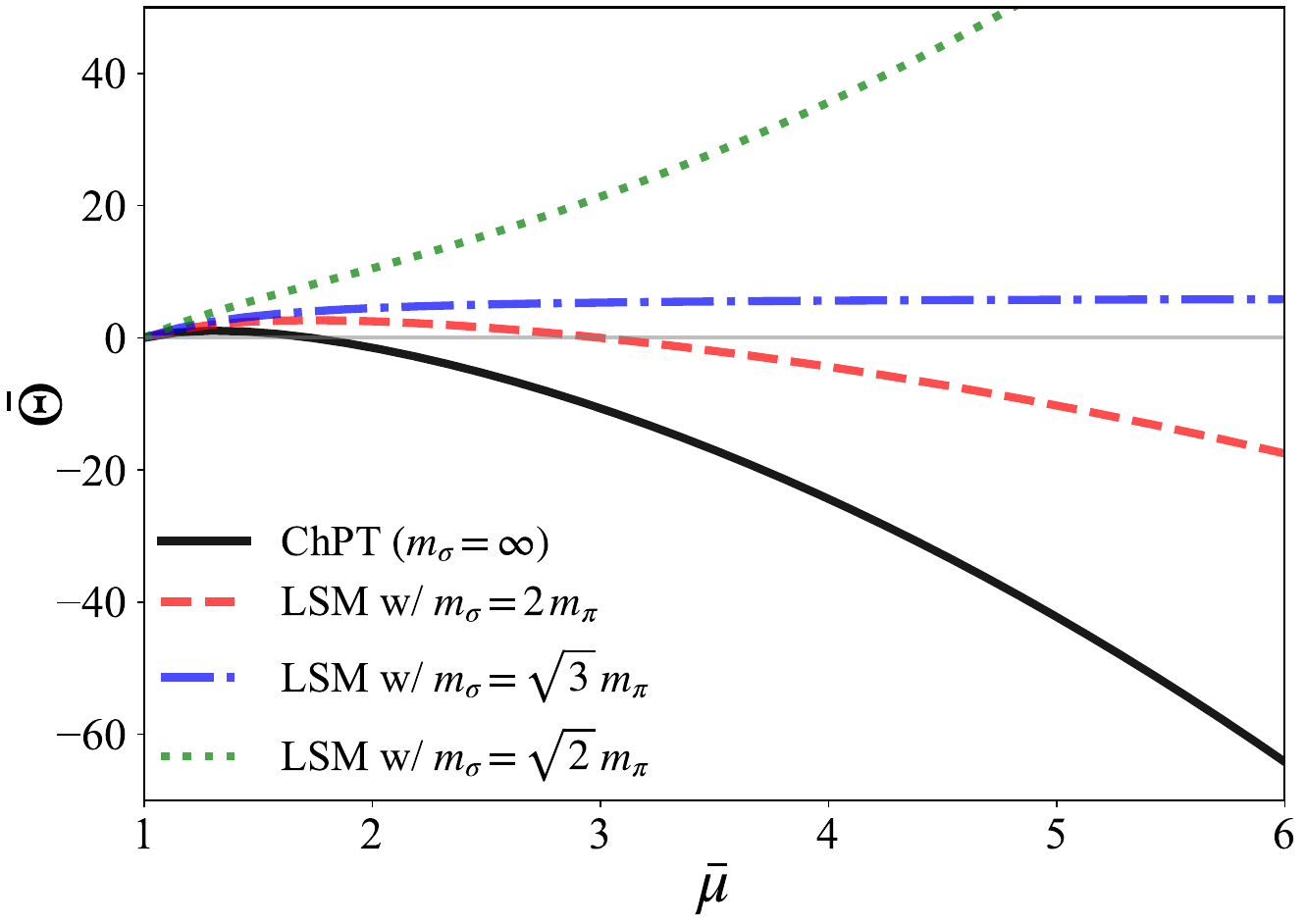}
\end{center}
\caption{
$\mu_q$ dependence on the normalized trace anomaly $\bar \Theta$.
}
\label{trace_ano}
\end{figure}

In order to visualize those behaviors of the trace anomaly, we depict $\mu_q$ dependence of $\bar{\Theta}$ in Fig.~\ref{trace_ano}. From this figure,
it is evident that 
the finite $\sigma$ meson mass for $m_\sigma>\sqrt{3}m_\pi$ 
plays the role in yielding the negative trace anomaly which has indeed been observed on the lattice simulation.



\section{Summary and discussion}
\label{sec:Conclusions}

In this paper, we have investigated the sound velocity at the finite quark chemical potential and zero temperature in QC$_2$D with two quark flavors, based on an effective model described by hadronic degrees of freedom. In particular, we have employed the linear sigma model, LSM, invented in Ref.~\cite{Suenaga:2022uqn} that is capable of describing not only the NG bosons associated with the chiral symmetry breaking but also their chiral partners as higher excitations, to go beyond the leading-order ChPT framework~\cite{Kogut:1999iv,Kogut:2000ek}.

Utilizing a mean-field approximation, we have analytically found that the thermodynamic quantities evaluated within the LSM are expressed by sums of the ChPT results and the corrections. Notably, the latter corrections are characterized by the inverse of a mass difference between the chiral partners, the $\sigma$ meson and pion. Hence, in a heavy mass limit of the $\sigma$ meson, the corrections vanish and the thermodynamic quantities in our LSM are reduced to the ChPT results, as naively expected. For this reason, in the QC$_2$D thermodynamic aspects, the LSM is regarded as a reasonable extension of the ChPT motivated by chiral symmetry. Therefore, we conclude that the LSM evaluation of the QC$_2$D thermodynamics can serve as a new benchmark of hadron model approaches to compare with lattice observations.

Within the ChPT analysis, the sound velocity in a sufficiently dense region converges on $c_s^2\to1$, meanwhile, based on the LSM, we have found that 
the asymptotic behavior of the sound velocity is affected by the $\sigma$ meson mass to approach the conformal limit $c_s^2\to1/3$. In addition, the $\sigma$ meson contribution provides only one peak in the sound velocity when $m_\sigma>\sqrt{3} m_\pi$ is satisfied. Furthermore, this peak has been found to emerge only for $ \mu_q>2\mu_q^{\rm cr}$, where $\mu_q^{\rm cr}=m_\pi/2$ denotes the critical chemical potential to enter the baryon superfluid phase.

We have further investigated the correlation between the sound velocity peak and the sign of the trace anomaly within the LSM. When the sound velocity peak appears, the sign of the trace anomaly turns negative with the increase of the chemical potential. Conversely, when the sound velocity peak is absent, the trace anomaly becomes always positive. We note that the QC$_2$D lattice simulation indicated both the presence of the sound velocity peak and the negative trace anomaly at some chemical potential~\cite{Iida:2022hyy}.

We expect that the revealed features of the sound velocity and trace anomaly provide useful information on roles of hadronic degrees of freedom in cold dense QC$_2$D from the chiral partner aspects, and predictions for future QC$_2$D lattice simulations. 

In what follows, we give comments and discussions that are not addressed in this paper. In this work, we have employed the LSM to incorporate the chiral partners as excitations that are not inherent in the ChPT, toward examination of, particularly, the peak structure of the sound velocity. For another direction to go beyond the lowest-energy regime of QC$_2$D, one would consider higher-order corrections within the ChPT framework. In fact, the ChPT Lagrangian including ${\cal O}(p^4)$ as well as ${\cal O}(p^2)$ terms at the mean-field level is also capable of yielding the sound velocity peak despite containing unfixed parameters. It is well known that the ChPT is useful to explain the low-energy NG-boson dynamics because of its systematic treatment. Meanwhile, the LSM has a great advantage that it can describe the chiral partners, i.e., $P$-wave excited hadrons, in addition to the NG bosons respecting chiral symmetry, and lattice simulations indeed observed such excited hadrons~\cite{Murakami:2022lmq}. Besides, there exits an argument that NG-boson contributions in all orders of the momentum expansion within the ChPT would be replaced by a $\sigma$ meson dynamics, in terms of the pion scattering lengths~\cite{Black:2009bi}. For these reasonings, our present LSM analysis is expected to be a promising approach to extend the ChPT results on the thermodynamics in cold dense QC$_2$D.

Our findings are evaluated within the mean-field approximation. To confirm the validity of our evaluations at the quantum level, it is worth going beyond the mean-field approximation in the LSM. Indeed, 
the one-loop calculation of the NG-boson fluctuations
has already been taken into account in the ChPT analysis~\cite{Splittorff:2001fy,Splittorff:2002xn,Andersen:2023ivj}. Given this fact, the inclusion of the chiral partner contribution in the one-loop calculation would be straightforward. More explicit analysis of such loop corrections in the LSM
is to be pursued elsewhere. 

Our present investigation has been mostly devoted to examining the sound velocity in the hadronic level, utilizing the LSM treating spin-$0$ hadrons. As the chemical potential is increased, it is expected that further excitations such as spin-$1$ hadrons would also start to contribute to the sound velocity~\cite{Suenaga:2023xwa}, and finally the quark degrees of freedom appear. 
However, how those other excited hadrons have influences on the sound velocity and how the quark-gluon dynamics begins to manifest themselves in the baryonic matter remain unclear~\cite{McLerran:2018hbz,Kojo:2021wax,Fujimoto:2020tjc,Braun:2022jme,Chiba:2023ftg,Suenaga:2019jjv,Kojo:2021knn}, owing to the current limited lattice data in cold dense QC$_2$D. Moreover, the $\sigma$ meson spectrum may be contaminated by a $0^+$ glueball through their mixings. Hence, in order to gain insights into those contributions, we expect more lattice results on not only the sound velocity but also the hadron spectrum in the future. Despite those uncertainties, our main harvest, the reproduction of the sound velocity peak in cold dense QC$_2$D by extending the NG boson dynamics to include the chiral partners, would already have great significance.

\section*{ACKNOWLEDGMENT}
This work of M.K. is supported in part by the National Natural Science Foundation of China (NSFC) Grant  No. 12235016,  and the Strategic Priority Research Program of Chinese Academy of Sciences under Grant No XDB34030000. 
D.S. was supported by the RIKEN special postdoctoral researcher program and by the Japan Society for the Promotion of Science (JSPS) KAKENHI Grants No.~23K03377 and 23H05439.
The authors thank Naoki Yamamoto for fruitful discussions and comments and Etsuko Itou and Kei Iida for useful information on the lattice data of the sound velocity and trace anomaly.

\appendix
\section{
Spontaneous chiral symmetry breaking in low-energy QC$_2$D and LSM
}
\label{app}

In this appendix, we derive the connection between the chiral condensate and pion decay constant in Eq.~(\ref{sigma_decay_con}).

We start from the underlying Lagrangian of the two-flavor QC$_2$D, 
\begin{eqnarray}
{\cal L}_{\rm QC_2D}=
\bar \psi (i\gamma^\mu D_\mu-m_l)\psi 
- \frac{1}{4}G_{\mu\nu}^aG^{\mu\nu,a},
\label{QCD_lag_theta}
\end{eqnarray}
where $\psi=(u,d)^T$ denotes the two-flavor quark doublet, and $D_\mu\psi=(\partial_\mu+igA_\mu^aT_c^a)\psi$ is the covariant derivative incorporating interactions with a gluon field $A_\mu^a$. The $2\times2$ matrix $T_c^a=\tau_c^a/2$ is the generator of $SU(2)_c$ color group with $\tau_c^a$ being the Pauli matrix; $g$ and $m_l$ are the QC$_2$D coupling constant and an isospin symmetric mass of current quarks, $m_u=m_d\equiv m_l$.
For the sake of convenience in the following discussion,
we take the Weyl representation of
the Dirac matrices,
\begin{eqnarray}
\gamma^\mu&=&
\begin{pmatrix}
0&\sigma^\mu\\
\bar \sigma^\mu&0
\end{pmatrix},\;\;\;
\gamma_5=
\begin{pmatrix}
1&0\\
0& -1
\end{pmatrix},
\end{eqnarray}
where 
$\sigma^\mu = ({\bm 1},\sigma^i)$ and 
$\bar \sigma^\mu = ({\bm 1},-\sigma^i)$ with 
the Pauli matrix $\sigma^i$ in spinor space.
By using the Weyl representation,
the quark field, which belongs to the fundamental representation for $SU(2)_c$, is decomposed into the left- and right-handed components,
\begin{eqnarray}
\psi = (\psi_R,\psi_L)^T.
\end{eqnarray}

From the pseudoreal viewpoint of $SU(2)$ groups, 
it is convenient to introduce the following ``conjugate fields'',
\begin{eqnarray}
\tilde\psi_{L,R}&=&
\sigma_2 \tau_c^2 \psi^*_{L,R}\nonumber\\
&=& (\tilde u_{L,R}, \tilde d_{L,R})^T.
\end{eqnarray}
Note that this pseudoreal representation satisfies a property 
similar to the Grassmann property for fundamental quarks
$\psi_{L(R)}^T \psi_{L(R)}=0$,
\begin{eqnarray}
\tilde\psi_{L(R)}^\dagger \psi_{L(R)}=0.
\end{eqnarray}
By combining the fundamental and pseudoreal representations, 
one can define the following field,
\begin{eqnarray}
\Psi= (\psi_R, \tilde \psi_L)^T=(u_R,d_R, \tilde u_L,\tilde d_L)^T,
\end{eqnarray}
where these four components are in the flavor space.
Then, the two-flavor QC$_2$D Lagrangian can be expressed as 
\begin{eqnarray}
&&{\cal L}_{\rm QC_2D}=
\Psi^\dagger 
\sigma^\mu
\left(
i\partial_\mu 
-gA_\mu^aT_c^a
\right)
\Psi
- \frac{1}{4}G_{\mu\nu}^aG^{\mu\nu,a}
\nonumber\\
&&
\;\;\;\;\;\;\;\;\;\;\;\;
-\frac{m_l}{2}
\left(\Psi^T \sigma^2 \tau_c^2 E^T \Psi+
({\rm  H.c.})
\right)
\label{Lag_su4}
.
\end{eqnarray}
At the massless limit of the current quark masses, 
this Lagrangian is invariant under the $SU(4)$ transformation,
\begin{eqnarray}
\Psi \to g \Psi
\;\;
\mbox{with $g\in SU(4)$}.
\end{eqnarray}

By using the $\Psi$ field, the quark condensate can be  described as 
\begin{eqnarray}
\langle \bar \psi \psi \rangle= 
\langle 
\left(\frac{1}{2}\Psi^T \sigma^2 \tau_c^2 E^T \Psi+
({\rm  H.c.})
\right)
\rangle.
\label{condenate_SU4}
\end{eqnarray}
This quark condensate constructed by $\Psi$ 
is not invariant under the $SU(4)$ transformation but
invariant
under the $Sp(4)$ transformation, 
\begin{eqnarray}
\Psi \to h \Psi\;\;
\mbox{with $h\in Sp(4)$}.
\end{eqnarray}
Namely, 
when the quark condensate gets a finite value,
the $SU(4)$ symmetry is spontaneously broken into the $Sp(4)$ symmetry in the vacuum of the two-flavor QC$_2$D. 
Consequently, five NG bosons show up, which is associated with 
the number of the broken generators of $SU(4)/Sp(4)$, $X^i$ ($i=1 \sim 5$). 
Indeed, a current of the broken symmetry is coupled to
the NG bosons.

Under the broken symmetry transformation with the rotation angle $\theta^i$, 
the quark field $\Psi$ is transformed as
\begin{eqnarray}
\Psi\to \exp[-i\theta^i X^i ] \Psi.
\label{axial_tra}
\end{eqnarray}
Then, one can obtain the broken current,
\begin{eqnarray}
j_5^{i\mu}&=&
\Psi^\dagger\sigma_\mu X^i \Psi. 
\end{eqnarray}
Note that this current can be rewritten by the original field $\psi$,
\begin{eqnarray}
&&j_5^{i\mu}
=
\nonumber\\
&&
\begin{cases}
\frac{1}{2\sqrt{2}}\bar \psi \gamma^\mu \gamma_5 \tau^i_f\psi
\;\;
(\mbox{for }i=1,2,3)
\\
\frac{i}{4\sqrt{2}}
\psi^T C \gamma_5
\tau^2_c (D_f^i)^\dagger (\gamma^\mu)^\dagger \psi 
+
({\rm H.c.})
\;\;(\mbox{for }i=4,5)
\end{cases}
\nonumber\\
\end{eqnarray}
where $C= i\gamma^2 \gamma^0$ denotes the charge-conjugation operator.
As a consequence of the spontaneous chiral symmetry breaking,
there exists an overlap amplitude between
the broken current $j^{i\mu}_5$ and the NG bosons $\mathscr{P}^i$
in the low-energy regime of QC$_2$D,
\begin{eqnarray}
\langle 0| j^{i\mu}_5(x) | \mathscr{P}^j(p) \rangle = -
i f_\pi  p^\mu   e^{-ip\cdot x } \delta^{ij}
\;\;\;
(\mbox{$i,j=1 \sim 5$})
,
\nonumber\\
\label{sand_axial}
\end{eqnarray}
where $f_\pi$ denotes the pion decay constant.

In the LSM, we can also define the broken current at the vacuum ($\mu_q=0$).
By following the underlying transformation in Eq.~(\ref{axial_tra}), $\Sigma$ also transforms under the broken symmetry of $SU(4)$ as
\begin{eqnarray}
\Sigma \to \exp[-i\theta^i X^i ]  \Sigma \exp[-i\theta^i (X^i)^T ]. 
\end{eqnarray}
From Eq.~(\ref{LSM_lag}),
the broken current of the LSM reads 
\begin{eqnarray}
j_5^{i\mu}=
2
(\partial^\mu \mathscr{P}^j \mathscr{S}^k
-\partial^\mu \mathscr{S}^j  \mathscr{P}^k
  )
  {\rm tr}\Bigl[
\{X^j, X^k\}
X^i
\Bigl].
\end{eqnarray}
In the hadronic phase where the vacuum undergoes the spontaneous chiral symmetry breaking,
$\sigma_0= \langle \mathscr{S}^0 \rangle\ \neq 0 $ and $ \Delta= \langle \mathscr{P}^5 \rangle =0$, 
the broken current takes the form of
\begin{eqnarray}
j_5^{i\mu}
&=&
\frac{1}{\sqrt{2}}
\sigma_0
\partial^\mu \mathscr{P}^i
+\cdots.
\end{eqnarray}
Here we have picked up a term relevant to NG bosons.
Thus, within the LSM, the overlap
amplitude between the broken current $j_5^{i\mu}$ and the NG bosons $\mathscr{P}^i$ is evaluated as
\begin{eqnarray}
\langle 0| j^{i\mu}_5(x) | \mathscr{P}^j(p) \rangle =-
i \frac{1}{\sqrt{2}}
\sigma_0  p^\mu   e^{-ip\cdot x }\delta^{ij}
.
\end{eqnarray}
Matching to the underlying definition in Eq.~(\ref{sand_axial}), the chiral condensate can be connected to the pion decay constant as
\begin{eqnarray}
\sigma_0 = \sqrt{2} f_\pi.
\end{eqnarray}
Note that the factor $\sqrt{2}$ deviates from the well-known case of the three-color QCD with two flavors since the chiral $SU(2)$ symmetry is extended to the $SU(4)$ symmetry due to the pseudoreality of QC$_2$D.


\begin{thebibliography}{10}

\bibitem{Baym:2017whm}
Gordon Baym, Tetsuo Hatsuda, Toru Kojo, Philip~D. Powell, Yifan Song, and
  Tatsuyuki Takatsuka.
\newblock {From hadrons to quarks in neutron stars: a review}.
\newblock {\em Rept. Prog. Phys.}, 81(5):056902, 2018.

\bibitem{Huang:2022mqp}
Yong-Jia Huang, Luca Baiotti, Toru Kojo, Kentaro Takami, Hajime Sotani, Hajime
  Togashi, Tetsuo Hatsuda, Shigehiro Nagataki, and Yi-Zhong Fan.
\newblock {Merger and Postmerger of Binary Neutron Stars with a Quark-Hadron
  Crossover Equation of State}.
\newblock {\em Phys. Rev. Lett.}, 129(18):181101, 2022.

\bibitem{Aarts:2015tyj}
Gert Aarts.
\newblock {Introductory lectures on lattice QCD at nonzero baryon number}.
\newblock {\em J. Phys. Conf. Ser.}, 706(2):022004, 2016.

\bibitem{Nagata:2021ugx}
Keitaro Nagata.
\newblock {Finite-density lattice QCD and sign problem: Current status and open
  problems}.
\newblock {\em Prog. Part. Nucl. Phys.}, 127:103991, 2022.

\bibitem{Muroya:2003qs}
Shin Muroya, Atsushi Nakamura, Chiho Nonaka, and Tetsuya Takaishi.
\newblock {Lattice QCD at finite density: An Introductory review}.
\newblock {\em Prog. Theor. Phys.}, 110:615--668, 2003.

\bibitem{Kogut:1999iv}
J.~B. Kogut, Misha~A. Stephanov, and D.~Toublan.
\newblock {On two color QCD with baryon chemical potential}.
\newblock {\em Phys. Lett. B}, 464:183--191, 1999.

\bibitem{Kogut:2000ek}
J.~B. Kogut, Misha~A. Stephanov, D.~Toublan, J.~J.~M. Verbaarschot, and
  A.~Zhitnitsky.
\newblock {QCD - like theories at finite baryon density}.
\newblock {\em Nucl. Phys. B}, 582:477--513, 2000.

\bibitem{Hands:1999md}
Simon Hands, John~B. Kogut, Maria-Paola Lombardo, and Susan~E. Morrison.
\newblock {Symmetries and spectrum of SU(2) lattice gauge theory at finite
  chemical potential}.
\newblock {\em Nucl. Phys. B}, 558:327--346, 1999.

\bibitem{Kogut:2001na}
J.~B. Kogut, D.~K. Sinclair, S.~J. Hands, and S.~E. Morrison.
\newblock {Two color QCD at nonzero quark number density}.
\newblock {\em Phys. Rev. D}, 64:094505, 2001.

\bibitem{Hands:2001ee}
Simon Hands, Istvan Montvay, Luigi Scorzato, and Jonivar Skullerud.
\newblock {Diquark condensation in dense adjoint matter}.
\newblock {\em Eur. Phys. J. C}, 22:451--461, 2001.

\bibitem{Muroya:2002ry}
Shin Muroya, Atsushi Nakamura, and Chiho Nonaka.
\newblock {Behavior of hadrons at finite density: Lattice study of color SU(2)
  QCD}.
\newblock {\em Phys. Lett. B}, 551:305--310, 2003.

\bibitem{Chandrasekharan:2006tz}
Shailesh Chandrasekharan and Fu-Jiun Jiang.
\newblock {Phase-diagram of two-color lattice QCD in the chiral limit}.
\newblock {\em Phys. Rev. D}, 74:014506, 2006.

\bibitem{Hands:2006ve}
Simon Hands, Seyong Kim, and Jon-Ivar Skullerud.
\newblock {Deconfinement in dense 2-color QCD}.
\newblock {\em Eur. Phys. J. C}, 48:193, 2006.

\bibitem{Alles:2006ea}
B.~Alles, Massimo D'Elia, and M.~P. Lombardo.
\newblock {Behaviour of the topological susceptibility in two colour QCD across
  the finite density transition}.
\newblock {\em Nucl. Phys. B}, 752:124--139, 2006.

\bibitem{Hands:2007uc}
Simon Hands, Peter Sitch, and Jon-Ivar Skullerud.
\newblock {Hadron Spectrum in a Two-Colour Baryon-Rich Medium}.
\newblock {\em Phys. Lett. B}, 662:405--412, 2008.

\bibitem{Hands:2010gd}
Simon Hands, Seyong Kim, and Jon-Ivar Skullerud.
\newblock {A Quarkyonic Phase in Dense Two Color Matter?}
\newblock {\em Phys. Rev. D}, 81:091502, 2010.

\bibitem{Hands:2011hd}
Simon Hands and Philip Kenny.
\newblock {Topological Fluctuations in Dense Matter with Two Colors}.
\newblock {\em Phys. Lett. B}, 701:373--377, 2011.

\bibitem{Cotter:2012mb}
Seamus Cotter, Pietro Giudice, Simon Hands, and Jon-Ivar Skullerud.
\newblock {Towards the phase diagram of dense two-color matter}.
\newblock {\em Phys. Rev. D}, 87(3):034507, 2013.

\bibitem{Hands:2012yy}
Simon Hands, Seyong Kim, and Jon-Ivar Skullerud.
\newblock {Non-relativistic spectrum of two-color QCD at non-zero baryon
  density}.
\newblock {\em Phys. Lett. B}, 711:199--204, 2012.

\bibitem{Boz:2013rca}
Tamer Boz, Seamus Cotter, Leonard Fister, Dhagash Mehta, and Jon-Ivar
  Skullerud.
\newblock {Phase transitions and gluodynamics in 2-colour matter at high
  density}.
\newblock {\em Eur. Phys. J. A}, 49:87, 2013.

\bibitem{Braguta:2016cpw}
V.~V. Braguta, E.~M. Ilgenfritz, A.~Yu. Kotov, A.~V. Molochkov, and A.~A.
  Nikolaev.
\newblock {Study of the phase diagram of dense two-color QCD within lattice
  simulation}.
\newblock {\em Phys. Rev. D}, 94(11):114510, 2016.

\bibitem{Boz:2018crd}
Tamer Boz, Ouraman Hajizadeh, Axel Maas, and Jon-Ivar Skullerud.
\newblock {Finite-density gauge correlation functions in QC2D}.
\newblock {\em Phys. Rev. D}, 99(7):074514, 2019.

\bibitem{Astrakhantsev:2018uzd}
N.~Yu. Astrakhantsev, V.~G. Bornyakov, V.~V. Braguta, E.~M. Ilgenfritz, A.~Yu.
  Kotov, A.~A. Nikolaev, and A.~Rothkopf.
\newblock {Lattice study of static quark-antiquark interactions in dense quark
  matter}.
\newblock {\em JHEP}, 05:171, 2019.

\bibitem{Iida:2019rah}
Kei Iida, Etsuko Itou, and Tong-Gyu Lee.
\newblock {Two-colour QCD phases and the topology at low temperature and high
  density}.
\newblock {\em JHEP}, 01:181, 2020.

\bibitem{Wilhelm:2019fvp}
Jonas Wilhelm, Lukas Holicki, Dominik Smith, Bj\"orn Wellegehausen, and Lorenz
  von Smekal.
\newblock {Continuum Goldstone spectrum of two-color QCD at finite density with
  staggered quarks}.
\newblock {\em Phys. Rev. D}, 100(11):114507, 2019.

\bibitem{Boz:2019enj}
Tamer Boz, Pietro Giudice, Simon Hands, and Jon-Ivar Skullerud.
\newblock {Dense two-color QCD towards continuum and chiral limits}.
\newblock {\em Phys. Rev. D}, 101(7):074506, 2020.

\bibitem{Iida:2020emi}
Kei Iida, Etsuko Itou, and Tong-Gyu Lee.
\newblock {Relative scale setting for two-color QCD with $N_f$=2 Wilson
  fermions}.
\newblock {\em PTEP}, 2021(1):013B05, 2021.

\bibitem{Astrakhantsev:2020tdl}
N.~Astrakhantsev, V.~V. Braguta, E.~M. Ilgenfritz, A.~Yu. Kotov, and A.~A.
  Nikolaev.
\newblock {Lattice study of thermodynamic properties of dense QC$_2$D}.
\newblock {\em Phys. Rev. D}, 102(7):074507, 2020.

\bibitem{Bornyakov:2020kyz}
V.~G. Bornyakov, V.~V. Braguta, A.~A. Nikolaev, and R.~N. Rogalyov.
\newblock {Effects of Dense Quark Matter on Gluon Propagators in Lattice
  QC$_2$D}.
\newblock {\em Phys. Rev. D}, 102:114511, 2020.

\bibitem{Murakami:2022lmq}
Kotaro Murakami, Daiki Suenaga, Kei Iida, and Etsuko Itou.
\newblock {Measurement of hadron masses in 2-color finite density QCD}.
\newblock {\em PoS}, LATTICE2022:154, 2023.

\bibitem{Braguta:2023yhd}
Victor~V. Braguta.
\newblock {Phase Diagram of Dense Two-Color QCD at Low Temperatures}.
\newblock {\em Symmetry}, 15(7):1466, 2023.

\bibitem{Iida:2022hyy}
Kei Iida and Etsuko Itou.
\newblock {Velocity of sound beyond the high-density relativistic limit from
  lattice simulation of dense two-color QCD}.
\newblock {\em PTEP}, 2022(11):111B01, 2022.

\bibitem{Detmold:2012wc}
William Detmold, Kostas Orginos, and Zhifeng Shi.
\newblock {Lattice QCD at non-zero isospin chemical potential}.
\newblock {\em Phys. Rev. D}, 86:054507, 2012.

\bibitem{Brandt:2018bwq}
Bastian~B. Brandt, Gergely Endrodi, Eduardo~S. Fraga, Mauricio Hippert, Jurgen
  Schaffner-Bielich, and Sebastian Schmalzbauer.
\newblock {New class of compact stars: Pion stars}.
\newblock {\em Phys. Rev. D}, 98(9):094510, 2018.

\bibitem{Brandt:2022hwy}
Bastian~B. Brandt, Francesca Cuteri, and Gergely Endrodi.
\newblock {Equation of state and speed of sound of isospin-asymmetric QCD on
  the lattice}.
\newblock {\em JHEP}, 07:055, 2023.

\bibitem{Abbott:2023coj}
Ryan Abbott, William Detmold, Fernando Romero-L\'opez, Zohreh Davoudi, Marc
  Illa, Assumpta Parre\~no, Robert~J. Perry, Phiala~E. Shanahan, and Michael~L.
  Wagman.
\newblock {Lattice quantum chromodynamics at large isospin density}.
\newblock {\em Phys. Rev. D}, 108(11):114506, 2023.

\bibitem{Hippert:2024hum}
Mauricio Hippert, Jorge Noronha, and Paul Romatschke.
\newblock {Upper Bound on the Speed of Sound in Nuclear Matter from Transport}.
\newblock 2 2024.

\bibitem{Kojo:2021wax}
Toru Kojo, Gordon Baym, and Tetsuo Hatsuda.
\newblock {Implications of NICER for Neutron Star Matter: The QHC21 Equation of
  State}.
\newblock {\em Astrophys. J.}, 934(1):46, 2022.

\bibitem{Fujimoto:2020tjc}
Yuki Fujimoto and Kenji Fukushima.
\newblock {Equation of state of cold and dense QCD matter in resummed
  perturbation theory}.
\newblock {\em Phys. Rev. D}, 105(1):014025, 2022.

\bibitem{Braun:2022jme}
Jens Braun, Andreas Gei\ss{}el, and Benedikt Schallmo.
\newblock {Speed of sound in dense strong-interaction matter}.
\newblock 6 2022.

\bibitem{McLerran:2018hbz}
Larry McLerran and Sanjay Reddy.
\newblock {Quarkyonic Matter and Neutron Stars}.
\newblock {\em Phys. Rev. Lett.}, 122(12):122701, 2019.

\bibitem{Kojo:2021ugu}
Toru Kojo.
\newblock {Stiffening of matter in quark-hadron continuity}.
\newblock {\em Phys. Rev. D}, 104(7):074005, 2021.

\bibitem{Kojo:2021hqh}
Toru Kojo and Daiki Suenaga.
\newblock {Peaks of sound velocity in two color dense QCD: Quark saturation
  effects and semishort range correlations}.
\newblock {\em Phys. Rev. D}, 105(7):076001, 2022.

\bibitem{Chiba:2023ftg}
Ryuji Chiba and Toru Kojo.
\newblock {Sound velocity peak and conformality in isospin QCD}.
\newblock 4 2023.

\bibitem{Masuda:2012ed}
Kota Masuda, Tetsuo Hatsuda, and Tatsuyuki Takatsuka.
\newblock {Hadron\textendash{}quark crossover and massive hybrid stars}.
\newblock {\em PTEP}, 2013(7):073D01, 2013.

\bibitem{Kojo:2014vja}
Toru Kojo and Gordon Baym.
\newblock {Color screening in cold quark matter}.
\newblock {\em Phys. Rev. D}, 89(12):125008, 2014.

\bibitem{Suenaga:2019jjv}
Daiki Suenaga and Toru Kojo.
\newblock {Gluon propagator in two-color dense QCD: Massive Yang-Mills approach
  at one-loop}.
\newblock {\em Phys. Rev. D}, 100(7):076017, 2019.

\bibitem{Kojo:2021knn}
Toru Kojo and Daiki Suenaga.
\newblock {Thermal quarks and gluon propagators in two-color dense QCD}.
\newblock {\em Phys. Rev. D}, 103(9):094008, 2021.

\bibitem{Son:2000by}
D.~T. Son and Misha~A. Stephanov.
\newblock {QCD at finite isospin density: From pion to quark - anti-quark
  condensation}.
\newblock {\em Phys. Atom. Nucl.}, 64:834--842, 2001.

\bibitem{Son:2000xc}
D.~T. Son and Misha~A. Stephanov.
\newblock {QCD at finite isospin density}.
\newblock {\em Phys. Rev. Lett.}, 86:592--595, 2001.

\bibitem{Kanazawa:2009ks}
Takuya Kanazawa, Tilo Wettig, and Naoki Yamamoto.
\newblock {Chiral Lagrangian and spectral sum rules for dense two-color QCD}.
\newblock {\em JHEP}, 08:003, 2009.

\bibitem{Lenaghan:2001sd}
J.~T. Lenaghan, F.~Sannino, and K.~Splittorff.
\newblock {The Superfluid and conformal phase transitions of two color QCD}.
\newblock {\em Phys. Rev. D}, 65:054002, 2002.

\bibitem{Metlitski:2005db}
Max~A. Metlitski and Ariel~R. Zhitnitsky.
\newblock {Theta-parameter in 2 color QCD at finite baryon and isospin
  density}.
\newblock {\em Nucl. Phys. B}, 731:309--334, 2005.

\bibitem{Harada:2010vy}
Masayasu Harada, Chiho Nonaka, and Tetsuro Yamaoka.
\newblock {Masses of vector bosons in two-color dense QCD based on the hidden
  local symmetry}.
\newblock {\em Phys. Rev. D}, 81:096003, 2010.

\bibitem{Suenaga:2022uqn}
Daiki Suenaga, Kotaro Murakami, Etsuko Itou, and Kei Iida.
\newblock {Probing the hadron mass spectrum in dense two-color QCD with the
  linear sigma model}.
\newblock {\em Phys. Rev. D}, 107(5):054001, 2023.

\bibitem{Minamikawa:2020jfj}
Takuya Minamikawa, Toru Kojo, and Masayasu Harada.
\newblock {Quark-hadron crossover equations of state for neutron stars:
  constraining the chiral invariant mass in a parity doublet model}.
\newblock {\em Phys. Rev. C}, 103(4):045205, 2021.

\bibitem{Kawaguchi:2023olk}
Mamiya Kawaguchi and Daiki Suenaga.
\newblock {Fate of the topological susceptibility in two-color dense QCD}.
\newblock {\em JHEP}, 08:189, 2023.

\bibitem{Black:2009bi}
Deirdre Black, Amir~H. Fariborz, Renata Jora, Nae~Woong Park, Joseph Schechter,
  and M.~Naeem~Shahid.
\newblock {Remark on pion scattering lengths}.
\newblock {\em Mod. Phys. Lett. A}, 24:2285--2289, 2009.

\bibitem{Splittorff:2001fy}
K.~Splittorff, D.~Toublan, and J.~J.~M. Verbaarschot.
\newblock {Diquark condensate in QCD with two colors at next-to-leading order}.
\newblock {\em Nucl. Phys. B}, 620:290--314, 2002.

\bibitem{Splittorff:2002xn}
K.~Splittorff, D.~Toublan, and J.~J.~M. Verbaarschot.
\newblock {Thermodynamics of chiral symmetry at low densities}.
\newblock {\em Nucl. Phys. B}, 639:524--548, 2002.

\bibitem{Andersen:2023ivj}
Jens~O. Andersen, Qing Yu, and Hua Zhou.
\newblock {Pion condensation in QCD at finite isospin density, the dilute Bose
  gas, and speedy Goldstone bosons}.
\newblock {\em Phys. Rev. D}, 109(3):034022, 2024.

\bibitem{Suenaga:2023xwa}
Daiki Suenaga, Kotaro Murakami, Etsuko Itou, and Kei Iida.
\newblock {Mass spectrum of spin-one hadrons in dense two-color QCD: Novel
  predictions by extended linear sigma model}.
\newblock 12 2023.

\end{thebibliography}
\end{document}